\documentclass[10pt,letterpaper]{article}
\usepackage[top=0.85in,left=1.5in,footskip=0.75in]{geometry}

\usepackage{graphicx}

\usepackage{subcaption}

\usepackage{amsmath,amssymb}

\usepackage{changepage}

\usepackage{textcomp,marvosym}

\usepackage{cite}

\usepackage{nameref,hyperref}

\usepackage[right]{lineno}

\usepackage[nopatch=eqnum]{microtype}
\DisableLigatures[f]{encoding = *, family = * }

\usepackage[table]{xcolor}

\usepackage{array}

\newcolumntype{+}{!{\vrule width 2pt}}

\newlength\savedwidth

\raggedright
\usepackage[aboveskip=1pt,labelfont=bf,labelsep=period,justification=raggedright,singlelinecheck=off]{caption}

\makeatletter
\renewcommand{\@biblabel}[1]{\quad#1.}
\makeatother

\usepackage{lastpage,fancyhdr,graphicx}
\usepackage{epstopdf}
\fancyheadoffset[L]{1.25in}
\fancyfootoffset[L]{1.25in}
\begin{document}
\vspace*{0.2in}

\begin{flushleft}

{\Large
\textbf\newline{Inhibition of bacterial growth by antibiotics: a minimal model} 
}
\newline
\\
Ledoux Barnabé\textsuperscript{1,*},
Lacoste David\textsuperscript{1}
\\
\bigskip
\textbf{1} Gulliver Laboratory, UMR CNRS 7083, PSL Research University, ESPCI, Paris F-75231, France

\bigskip

* barnabeledoux@gmail.com

\end{flushleft}

\section*{Abstract}
Growth in bacterial populations generally depends on the environment (availability and quality of nutrients, presence of a toxic inhibitor, product inhibition..). Here, we build a minimal model to describe the action of a bacteriostatic antibiotic, assuming that this drug inhibits an essential autocatalytic cycle involved in the cell metabolism. The model recovers known growth laws, can describe various types of antibiotics and confirms the existence of two distinct regimes of growth-dependent susceptibility, previously identified only for ribosome targeting antibiotics. We introduce a proxy for cell risk, which proves useful to compare the effects of various types of antibiotics. We also develop extensions of our model to describe the effect of combining two antibiotics targeting two different autocatalytic cycles or a regime where cell growth is inhibited by a waste product.

\section*{Introduction}
The emergence of antibiotic resistance, which often occurs under changing levels of antibiotics is a major concern for human health \cite{davies_j_origins_2010,chait_antibiotic_2007}. 
In an important class of antibiotics, called bacteriostatic antibiotics \cite{loree_bacteriostatic_2023}, the drug does not induce death directly, but renders some essential process in the cell metabolism less efficient or inactive \cite{lin_ribosome-targeting_2018,contreras_binding_1974,mondal_impact_2014,mosaei_mechanisms_2019,kohanski_mistranslation_2008,si_invariance_2017} resulting in a reduced cell growth. For these antibiotics, it is essential to properly model cell metabolism and cell growth in order to better understand the action of antibiotics \cite{levin_numbers_2017,tuomanen_rate_1986,lopatkin_bacterial_2019,greulich_growth-dependent_2015}. According to  \cite{svetlov_kinetics_2017}, the distinction between bacteriostatic and bactericidal antibiotics that target ribosomes depends on the value of their dissociation rate from the ribosomes. Antibiotics with slow dissociation rates are more likely to induce cell death and be classified instead as bactericidal because of the depletion of essential proteins they cause.

In the field of bacterial growth, the study of growth laws  \cite{scott_bacterial_2011,wu_cellular_2022,scott_interdependence_2010} represents a major step forward in our understanding of cell growth. These growth laws result from conservation of ribosome capacity and flux balance at steady-state. Recently, a new way to understand them has been put forward, that relies on a description of the cell metabolism as an ensemble of autocatalytic cycles, such as the cycle of ribosome production and that of RNA polymerase production \cite{roy_unifying_2021}. This method, which we will apply in this paper has wide applications for cell biology. For instance, it has been recently used to formulate predictions about the interplay between cellular growth rate and mRNA abundances \cite{calabrese_how_2024}. 

While some predictions about the action of RNA-polymerase targeting antibiotics have been derived in Ref. \cite{roy_unifying_2021}, the full consequences for the inhibition of growth by a general antibiotics have not. In particular, this work does not discuss the second growth law that describes the inhibition of translation by antibiotics, nor the possibility that there may be two different regimes for the action of antibiotics, namely the so-called reversible and irreversible binding regimes of antibiotics. 
This distinction is quite important in practice because for reversible binding, faster growth in the absence of the drug leads to an increased susceptibility, while the opposite is true for irreversible binding \cite{greulich_growth-dependent_2015}. Further, a regime of antibiotics concentration exists where two values of growth rate are possible (growth rate heterogeneity or bistability \cite{deris_innate_2013}) for the same range of physical parameters. At the moment, it is not known whether this behavior should be expected for all types of antibiotics.

To summarize, we believe that the inhibition of bacterial growth by antibiotics has not been considered from a sufficiently general point of view, which motivates the present study. 
By building on Refs. \cite{roy_unifying_2021} and \cite{greulich_growth-dependent_2015,allen_antibiotic_2016}, we develop a general framework for the inhibition of bacterial growth by bacteriostatic antibiotics in which the cell metabolism is modeled as coupled autocatalytic cycles. 
Given the central role played by Ref. \cite{greulich_growth-dependent_2015,allen_antibiotic_2016} in our work, we start by a quick summary of the main findings of this paper. In the next section, we present our model, so that the new elements which we have introduced appear clearly. Then, we explore some consequences, concerning growth laws, and we test our model with experimental data on the dependence of the growth rate as function of the concentration of antibiotics for a wide range of different antibiotics. Then, we introduce a new proxy of cell risk induced by the antibiotics and we present some extensions of our model for more complex situations involving the combined action of multiple antibiotics \cite{kavcic_minimal_2021,kavcic_mechanisms_2020,bollenbach_nonoptimal_2009} or the indirect  effect due to the accumulation of a product with inhibition properties.

\section*{Model for the inhibition of bacterial growth by antibiotics \label{sec:Greulich}} 
Here, we recapitulate the main findings of a classic model of inhibition of bacterial growth by antibiotics \cite{greulich_growth-dependent_2015}, which is applicable to antibiotics that target ribosomes. In this model, the cell is viewed as a compartment in which the antibiotic present outside the cell can enter and bind to ribosomes. The perturbation of translation produced by the antibiotics is described by growth laws, which quantify the interdependence of the cell growth rate $\lambda$ with the intracellular ribosome concentration $r$. 
The first law states that the ribosome concentration should increase linearly with the growth rate according to \cite{scott_bacterial_2011,wu_cellular_2022,scott_interdependence_2010}:
\begin{equation}
\label{GL_1}
    r_u= r_{min} + \frac{\lambda}{\kappa_t},
\end{equation}
where $\kappa_t$, $r_u$ and $r_{min}$ are respectively 
the translation capacity, the concentration of unperturbed ribosomes and a minimal ribosome concentration.

The second growth law states that, in the presence of an antibiotic inhibiting translation, the ribosome production is up-regulated, which also leads to another linear relation
\begin{equation}
\label{GL_2}
    r_{tot}= r_u+r_b = r_{max} - \lambda \Delta r \left( \frac{1}{\lambda_0} - \frac{1}{\kappa_t \Delta r} \right),
\end{equation}
where where $r_b$ is the concentration of ribosomes bound to the antibiotics and $\Delta r= r_{max}-r_{min}$ is the dynamic range of the ribosome concentration.
In other words, the second growth law describes the increased production of ribosomes that follows translation inhibition. As a result, the total ribosome concentration becomes negatively correlated with the bacterial growth rate in the presence of these inhibitors.

 Now the antibiotics enter the cell and bind to the ribosomes, with the rate $f(r_u,r_b,a)=-k_{on} a (r_u - r_{min})+k_{off} r_b$ where $k_{off}$ and $k_{on}$ are first and second order rate constants and $a$ is the antibiotic concentration inside the cell. Only ribosomes above the minimum threshold $r_{min}$ can bind to ribosomes according to this formula. The flux of antibiotic concentration into the cell is $J(a_{ex},a)= P_{in} a_{ex} - P_{out} a$, where $a_{ext}$ is the antibiotic concentration outside the cell. Antibiotics enter the cell with rate $P_{in}$ and exit with rate $P_{out}$, which could occur due to diffusion by passive transport or through pores by active transport \cite{chopra_molecular_1988, macnair_role_2023}.  
 
 Together, these assumptions lead to the following dynamical equations \cite{greulich_growth-dependent_2015}:
 \begin{align}
 \label{Greulich_model}
    \frac{da}{dt} = & - \lambda a + f(r_u,r_b,a) + J(a_{ex},a), \\
   \frac{dr_u}{dt}=   &  - \lambda r_u + f(r_u,r_b,a) + s(\lambda), \nonumber \\
   \frac{dr_b}{dt}= & -\lambda r_b - f(r_u,r_b,a), \nonumber
\end{align}
where $s(\lambda)=\lambda r_{tot}$ represents the ribosome synthesis rate.

In the absence of inhibitors, the pre-exposure or basal growth rate is $\lambda_0$, which corresponds to the normal behavior of the cell. 
The steady-state solution of this model is given by the following cubic equation \cite{greulich_growth-dependent_2015}:
\begin{equation}
\label{cubic_eq}
 0= \left( \frac{\lambda}{\lambda_0} \right)^3 - \left( \frac{\lambda}{\lambda_0} \right)^2+ \frac{\lambda}{\lambda_0} \left[ \frac{1}{4} \left( \frac{\lambda_0^*}{\lambda_0} \right) ^2 + \frac{a_{ex}}{2 IC_{50}^*} \frac{\lambda_0^*}{\lambda_0} \right] - \frac{1}{4} \left( \frac{\lambda_0^*}{\lambda_0} \right)^2.   
\end{equation}
The reversibility of the binding of the antibiotic is characterized by the parameter $\lambda_0^*=2 \sqrt{P_{out} K_D \lambda_0}$, $K_D$ is the dissociation constant $k_{off}/k_{on}$ and $IC_{50}^*$ is a typical concentration such that $IC_{50}^*=\Delta r \lambda_0^*/2P_{in}$.
Since Eq. \ref{cubic_eq} is a cubic equation in the growth rate, there are one or three solutions, and in particular there is a parameter regime in which the dynamical system can be multi-valued.

The model predicts two regimes depending on the value of $\lambda_0^*$, called the reversible and irreversible limits. The reversible limit $\lambda \ll \lambda_0^*$ describes a regime of strong outflux of toxic agents and unbinding rate. In that case, the growth rate has a smooth behavior described by:
\begin{equation}
\label{Greulich_reversible}
    \frac{\lambda}{\lambda_0} = \frac{1}{1+\frac{a_{ex}}{IC_{50}}}.
\end{equation}
This smooth behavior is due physically to a rapid equilibrium which is reached between intra and extra cellular antibiotic pools.

In contrast, the irreversible limit $\lambda \gg \lambda_0^*$ corresponds to negligible outflux and unbinding rate compared to the influx of toxic agents and binding rate. 
In that case, one obtains a discontinuous function: 
\begin{equation}
\label{Greulich_irreversible}
    \lambda=\frac{\lambda_0}{2}\left(1+\sqrt{1-\frac{4P_{in}a_{ex}}{\lambda_0}}\right).
\end{equation}
In this regime, the system behaves as a toggle switch behavior, due to the competition between the antibiotic influx and the ribosome production.

By analyzing various types of antibiotics, the authors of \cite{greulich_growth-dependent_2015} found that experimental data for bacteriostatic antibiotics indeed fit into one class or the other. Another major insight of the model, was the prediction of different growth dependent susceptibility for the two classes of antibiotic behaviors. 
This susceptibility is measured thanks to the half-inhibition concentration $IC_{50}$, which is defined as the concentration of toxic agent at which the growth rate is half its initial value. This is a measure of the sensitivity of the system to external stress, the higher it is, the more resistant is the system to inhibitors. 
By substituting $a_{ex}=IC_{50}$ and $\lambda=\lambda_0/2$ into Eq. \ref{cubic_eq}, one finds that the half inhibitory concentration $IC_{50}$ falls onto a universal growth dependent susceptibility curve: 
\begin{equation}
\label{Greulich_universal}
 \frac{IC_{50}}{IC_{50}^*}\simeq\frac{1}{2}\left(\frac{\lambda_0^*}{\lambda_0}+\frac{\lambda_0}{\lambda_0^*}\right). 
\end{equation}

\section*{Modified model based on autocatalytic cycles} 
\label{sec:math}
We now introduce our model for cell metabolism as two coupled 
autocatalytic cycles, in which one cycle describes the production of ribosomes, while the other describes RNA-polymerase production \cite{roy_unifying_2021}. These two autocatalytic cycles are coupled because ribosomes are necessary to synthesize RNA-polymerase protein subunits and vice-versa for ribosomes: $B_1$ represents the number of active ribosomes; $C_1$  the number of active RNA polymerases; similarly $B_2, ..., B_{N-1}$ and $C_2, ..., C_{K-1}$ are the abundances of intermediates involved in the assembly of ribosomes and RNA polymerases respectively, $B_N$; $C_K$ are  the abundances of fully assembled but resting ribosomes/RNA polymerases respectively.
Similarly to the classic model we first presented, we suppose that "toxic" inhibiting agents in numbers $A$ can bind to one of the autocatalysts (chosen here to be $B_1$ for simplicity) with a rate $k_{on}$ and unbind with a rate $k_{off}$, proportionally to the relative abundance of antibiotics in the cell \cite{greulich_growth-dependent_2015}. We denote $B_{1,u}$ the abundance of unbound ribosomes and $B_{1,b}$ the abundance of bound ribosomes.

The signification of the different variables in the model is 
summarized in the table \ref{tab:table_networks}.
\begin{table}[h]
    \centering
    \begin{tabular}{||c|c||}
         \hline $B_{1u}$ & Number of fully formed free active ribosomes \\
         \hline $B_{1b}$ & Number of fully formed ribosomes bound to antibiotics \\
         \hline $A$ & Number of toxic agent molecules within the cell \\
         \hline $a_{ex}$ & Concentration of toxic agent molecules outside the cell \\
        \hline $\Omega$ & Cell volume \\
         \hline $B_k$ for $k \ge 2$ & Number of ribosomes precursors \\
         \hline $C_1$ & Number of fully formed and active RNA-polymerases \\
         \hline $C_k$ for $k \ge 2$ & Number of RNA-polymerase precursors \\ 
         \hline $N$ (resp. $K$) & Number of building steps for ribosomes (resp. RNA-polymerase) \\ \hline
    \end{tabular}
    \caption{Variables of the model. Note that we used dimensionless numbers for species within the cell, except for $a_{ex}$ which has the unit of a concentration and $\Omega$ which has the unit of a volume.}
    \label{tab:table_networks}
\end{table}

In our model, we rely on Leontief's production function \cite{yamagishi_microeconomics_2021} (see Supplementary Material \cite{SM} Section E for details), according to which the rates of reactions involving two complementary resources are set by the limiting quantity among the two using a minimum function.
Historically, this law of the minimum has been introduced by Leontief's in his work in economy \cite{dobos_dynamic_2005}: in a network of firms producing one product each by consuming the outputs of other firms (resources), the rate of production will be set by the availability of the scarcer resource. A similar idea was developed later by Liebig in ecology \cite{oneill_multiple_1989}. More recently, it was used for modeling autocatalytic cycles in metabolism  \cite{roy_unifying_2021}.
With this method, we get linear equations in regimes where one reactant is scarce. This is similar to assuming that one reactant is in excess in a chemical reaction, and that the kinetics is set by the concentration of the scarcer reactant. 

\begin{figure*}
    \centering
    \includegraphics[width=\textwidth]{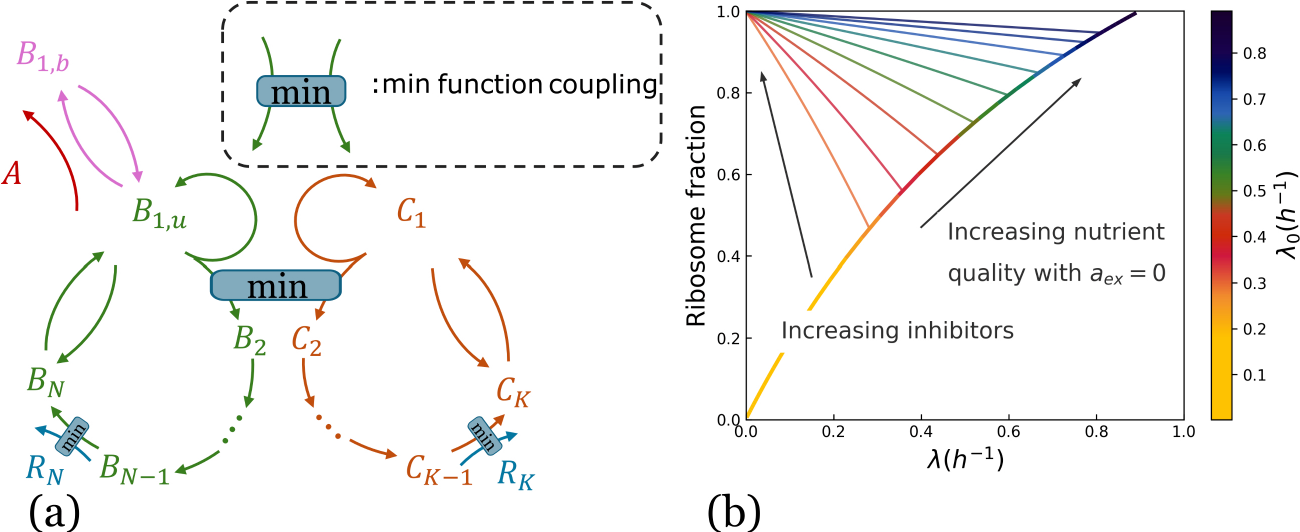}
    \caption{(a) Scheme of coupled autocatalytic networks interacting with a toxic agent. The blue box linking two arrows represents a coupling through a min function \cite{dobos_dynamic_2005,oneill_multiple_1989}. (b) The first growth law is the increase of the ribosome fraction with the growth rate (solid curve), the second law corresponds to the colored lines obtained by varying the amount of antibiotics. The pre-exposure growth rate $\lambda_0$ displayed on the right scale.}
    \label{fig:autocat_networks_antib}
\end{figure*}

Unlike in the previous model, which was formulated in terms of concentrations, our approach uses abundances or numbers \cite{pandey_analytic_2016} precisely because it is based on the Leontief framework. 
Due to this difference, our dynamical equations formulated in terms of species numbers do not contain the growth rate explicitly unlike Eq. \ref{Greulich_model}. Naturally, it is straightforward to show that the two formulations are equivalent, because of the assumption of balanced growth for the cell. In this regime, all species present in the autocatalytic cycles grow 
at the same rate $\lambda=d\ln{ {\mathcal N}}/dt$, where $\mathcal N$ is typically the number of ribosomes or RNA-polymerases... 
Note that the cell volume is not constant but also grows at the same rate as the abundances of species inside the cell. This is the reason for the use of fractions defined with respect to the total abundances of mature molecules $B_{tot}=B_{1,u}+B_{1,b}+B_{N}$. 

One can then combine the equations of the model to obtain a linear matrix equation for the sub-populations of ribosomes only, without explicit dependence on antibiotics,
and a self consistent equation for the growth rate $\lambda$ of the whole cycle (see Supplementary material \cite{SM}, section A). 
In the following, we assume the cycle targeted by the toxic agent becomes limiting. The effect of the inhibition of one cycle on the other cycle is only considered in Appendix, section \ref{ap:consequences}. Consequently, we isolate the inhibited cycle and study its growth, because it restricts the growth of the rest of the network.

\section*{Growth laws}
A key quantity is the fraction of active ribosomes $Q(\lambda) = B_{1,u} / B_{tot}$ ,
which takes the form of a polynomial in terms of the cell growth rate $\lambda$ with factors depending on the rate constants $k_{B,i}$ of reaction steps of the autocatalytic cycle :
\begin{equation}
    Q(\lambda)  =  \frac{1}{k_{B,1}}\left(1+\frac{\lambda}{k_{B,2}}\right)\times \hdots \times \left(1+\frac{\lambda}{k_{B,N-1}}\right)\left(\lambda + \frac{1}{\tau_{life}}\right),
\label{eq:Qlambda}
\end{equation}
where $\tau_{life}$ is the life time of mature intermediates $B_N$, $B_{1,u}$ and $B_{1,b}$,
which corresponds to the time of degradation of these molecules. This life time is assumed to be of the same order for all these species for simplicity and is typically large in comparison with the growth rate, unless the cell is in a regime of reduced growth \cite{calabrese_how_2024}. 

The expression above simplifies to $Q(\lambda) \simeq \lambda/ k_{B1}$ in the limit of "fast assembly" $k_{B,2}, ..., k_{B,N-1} \gg \lambda$ and long ribosome lifetime $\lambda\gg1/\tau_{life}$. In this case, the results do not depend on the number of steps $N$ in the first cycle. We also understand from Eq. \ref{eq:Qlambda} that if one step $n$ becomes limiting, the term $\lambda/k_{B,n}$ cannot be ignored, which modifies $Q(\lambda)$. With the above conditions, we recover the linear increase of the fraction of unbound ribosomes with respect to $\lambda$, which is the {\it first growth law}:
\begin{equation}
    \frac{B_{1,u}}{B_{tot}} \simeq \frac{\lambda}{k_{B,1}}+\frac{1}{k_{B,1}\tau_{life}}.
\end{equation}
Note that this is the equivalent of Eq. \ref{GL_1} in the previous model. This law describes the increase of the fraction of unbound ribosomes with the growth rate under changes of nutrient quality in the absence of antibiotics, so when $a_{ex}=0$. Here, increasing the nutrient quality can be realized by increasing assembly rates $k_{B,2}, ..., k_{B,N-1}$, assuming that they are equal to each other. Indeed, if only one of these rates were increased, the other steps would be limiting and we would not see the effect we are interested in. In the end, we obtain the solid blue curve in Fig. \ref{fig:autocat_networks_antib}b, which approaches the origin when $\lambda$ goes to zero due 
to the long lifetime assumption.

When an antibiotic inhibiting translation is present, the ribosome fraction $(B_{1,u}+B_{1,b})/B_{tot}$
decreases with the growth rate, which is the \textit{second growth law} \cite{scott_bacterial_2011}. With our formalism, we indeed obtain a negative correlation between these variables, which takes a linear form:
\begin{equation}   \frac{B_{1,u}+B_{1,b}}{B_{tot}} \simeq 1-\frac{\lambda}{k_{B,3}},
\label{eq:second_law}
\end{equation}
if we assume fast assembly, fast activation, long ribosome lifetime $\lambda\gg 1/\tau_{life}$ and a single intermediate step ($N=3$).
This equation is the equivalent of the second growth law described by Eq. \ref{GL_2}.
Without specific assumptions on the rates, one obtains the colored curves in  Fig.\ref{fig:autocat_networks_antib}b, which have been obtained by varying the external concentration of antibiotics $a_{ex}$ keeping all other parameters fixed. 

As the concentration of antibiotics increases, the growth rate always decreases below the basal growth rate $\lambda_0$.
We find that for $a_{ex}=0$, the decreasing and increasing curves of Fig.\ref{fig:autocat_networks_antib}b cross each other, which is expected because $\lambda = \lambda_0$ at this point. 

It is important to appreciate that the first and the second growth laws are derived from our model, while they were introduced as phenomenological constraints in Eq. \ref{GL_1} and Eq. \ref{GL_2} in the model we first presented. Further, in the original work on growth laws \cite{scott_interdependence_2010}, linear dependencies with respect to the growth rate were reported.
In contrast to this, we see from Fig. \ref{fig:autocat_networks_antib}, that neither the first nor the second growth law is strictly described by linear relations. In fact, a curvature can often be spotted in some works in the literature without their authors commenting about it. For instance, the solid line describing growth laws is clearly curved in the predictions from the complex stochastic cell model developed in Ref. \cite{thomas_sources_2018}. Thus, we can conclude that the observed curvature in the growth laws is predicted by theory and is not related to the stochasticity of biochemical reactions since it is already present in our minimal deterministic model.

\label{subsec:simple_case}
We now explore further consequences of our formalism. 
Let us first consider the case of arbitrary number of intermediates ($N$), for which we have obtained the self-consistent equation for $Q(\lambda)$ given in Eq. \ref{eq:self-consist}.

The reversible limit $\lambda \ll \lambda_0^*$ describes a regime of strong outflux of toxic agents and unbinding rate. We find that in this limit (see Supplementary material \cite{SM}, section A):
\begin{equation}
    Q(\lambda) = \frac{1}{1+\frac{K_D P_{in}}{P_{out}}a_{ex}}.
\end{equation}

In contrast, the irreversible limit $\lambda \gg \lambda_0^*$ corresponds to negligible outflux and unbinding rate compared to the influx of toxic agents and binding rate. Then, we obtain a different equation setting the growth rate  (see Supplementary material \cite{SM}, section A):
\begin{equation}
    Q(\lambda) = 1 + \frac{P_{in} a_{ex}}{\lambda}.
\end{equation}

For ribosomes in the regime of intermediate or high growth rates, we can expect a long lifetime, a small resting rate, fast assembly and fast activation  \cite{roy_unifying_2021}. These conditions translate to $\frac{1}{\tau_{life}},k_{B4}\ll\lambda_0 ,  k_{B1} \ll k_{B2}, ..., k_{B,N}$, yielding
$\lambda_0 \simeq k_{B1}$. In this limit, we can simplify our self-consistent equation for the growth rate
\begin{equation}
    \frac{P_{in}a_{ex}}{\left(\frac{k_{B1}}{k_{on}}\frac{\lambda + P_{out}}{\lambda}+\frac{\lambda}{\lambda+k_{off}}\right)}\simeq\left(1-\frac{\lambda}{\lambda_0}\right)\left(\lambda+k_{off}\right),
\label{eq:self_cons_ribosomes}
\end{equation}
so that we recover the equation derived in \cite{greulich_growth-dependent_2015}. With the additional assumption of fast binding $\lambda_0 \ll k_{on}$, the possible values of the growth rate are roots of a polynomial, from which it is possible to recover the reversible and irreversible limits of antibiotics binding described previously. Further, we find in this limit $Q \simeq \lambda/\lambda_0$.


Interestingly, the self-consistent equation for the growth rate obtained within the autocatalytic framework (see Supplementary material \cite{SM}, section A) has two solutions in the irreversible limit with fast assembly, leading to two separate branches of solutions for $\lambda$. A first solution remains close to $0$, corresponding to a non-growing cell. A second one is larger but exists only until a given concentration of inhibitors is reached, above which the system jumps on the other branch, and the growth rate vanishes as shown in Fig. \ref{fig:ic50_m0}a. In experiments, in the irreversible case, the system usually starts from $\lambda_0$ and the growth rate decreases as the concentration of inhibitors increases, until the discontinuity where the growth rate jumps on the second branch and vanishes. This growth rate heterogeneity happens above a threshold in terms of the antibiotic concentration. Such a phenomenon has been predicted in other theoretical works \cite{greulich_growth-dependent_2015,elf_bistable_2006}, and it has also been observed experimentally \cite{deris_innate_2013, irwin_evidence_2010}. 

\section*{Experimental test of the model}
\label{sec:results}
We have tested our model on a number of antibiotics, for which experimental data can be found in the literature \cite{si_invariance_2017, greulich_growth-dependent_2015}: Chloramphenicol inhibits ribosome production by binding to ribosomes, preventing them from transcribing new proteins; 
Rifampicin targets RNA-polymerase by binding to RNA-polymerase \cite{mcclure_mechanism_1978,campbell_structural_2001};
 Kanamycin, Streptomycin, Chloramphenicol and Erythromycin target the ribosomal autocatalytic cycle \cite{chaturvedi_protein_2016,lin_ribosome-targeting_2018,mondal_impact_2014,kohanski_mistranslation_2008};
 and finally Triclosan targets the synthesis of fatty acids \cite{heath_mechanism_1999,mcmurry_triclosan_1998,escalada_triclosan_2005}, thus affecting the building of bacterial membranes \cite{roy_unifying_2021}. 
\begin{figure*}
    \centering  
    \includegraphics[width=\textwidth]{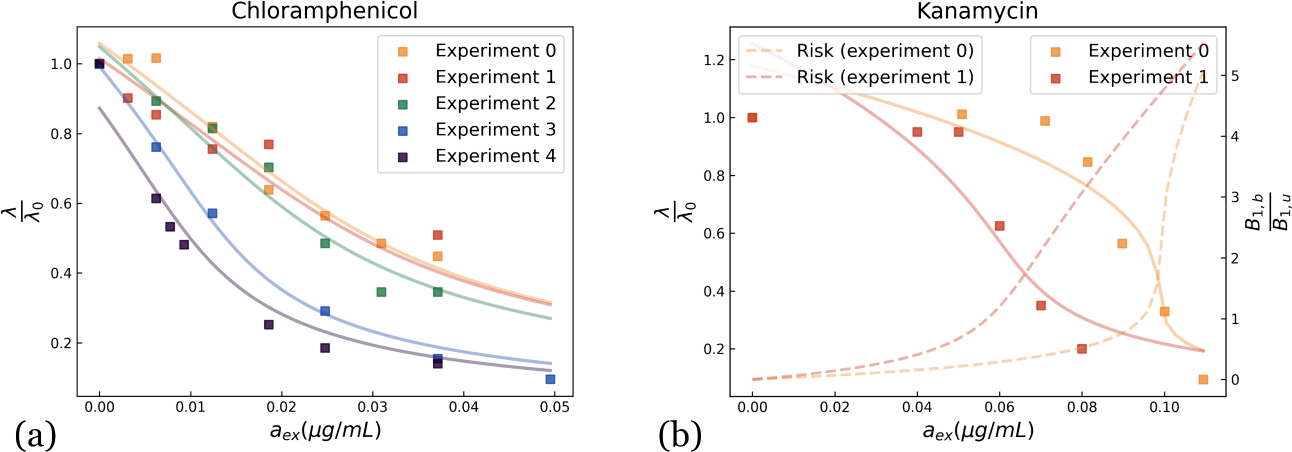}
    \caption{Comparison with experiments for two drugs affecting bacterial growth, namely (a) Chloramphenicol (data from \cite{si_invariance_2017} and \cite{greulich_growth-dependent_2015}) and (b) Kanamycin (data from \cite{greulich_growth-dependent_2015}). The solid line shows the growth rate as a function of the fraction of inhibitors, while the dotted line shows a measure of the risk faced by the cell defined in the text. The data were fitted by constraining the parameters as explained in Supplementary material \cite{SM}. Different experiments for the same antibiotic correspond to different growth medium.}
    \label{fig:compare_drugs}
\end{figure*}
In Fig.\ref{fig:compare_drugs}, we show the normalized growth rate $\lambda/\lambda_0$ as function of the concentration of antibiotics only for Chloramphenicol and Kanamycin, the plots for the other antibiotics are shown in Supplementary material \cite{SM}, section B.

In \cite{roy_unifying_2021}, the effects of Triclosan and Rifampicin were explained by adding Hill functions heuristically to describe saturation effects in the cycle. In contrast here, we provide an explicit expression for the dependence of the growth rate on the fraction of antibiotics without such an assumption. The fact that we are able to describe a large panel of antibiotics suggests that these antibiotics can indeed be depicted as inhibitors affecting essential cellular autocatalytic cycles despite their different mechanisms. Note that we recover different concavities in Fig.\ref{fig:compare_drugs}, which correspond to the two distinct regimes of cellular response to the antibiotics previously identified for ribosome-targeting antibiotics \cite{greulich_growth-dependent_2015}: the reversible limit where the outflux of antibiotics compensates the influx of the latter, and the irreversible limit where antibiotics bind quickly to autocatalysts, resulting in an accumulation of bound, inhibited individuals. 

\section*{Cell risk induced by the antibiotics}

Antibiotics have a rather limited number of targets such as ribosomes \cite{greulich_growth-dependent_2015,lin_ribosome-targeting_2018,mondal_impact_2014,loree_bacteriostatic_2023,levin_numbers_2017} or RNA-polymerase \cite{mosaei_mechanisms_2019} for instance. Regardless of the mechanism of action or precise targets, the effect of antibiotics on growth show similarities \cite{lopatkin_bacterial_2019}, which suggests that a general measure of the risk induced by the toxic agent might exist. 
In particular, bacteriocidal antibiotics do not appear to be fundamentally different from bacteriostatic ones, both reduce cell growth, but if the inhibition is too strong, processes that are necessary for survival cannot be satisfied and cell death can occur \cite{baquero_proximate_2021,levin_numbers_2017}. 
In fact, it has been demonstrated that for ribosome targeting antibiotics, the cidality depends on the rate of dissociation of antibiotics (and thus on the amount of bound antibiotics in the cell) \cite{svetlov_kinetics_2017}. This study concluded that cell death induced by a ribosome targeting drug results from a prolonged inhibition of synthesis. This means that for sufficiently slow dissociation rates, antibiotics stay bound to ribosomes. 

To quantify the effect of antibiotics, we introduce a measure of the risk faced by the cell, defined as the fraction of bound active individuals $B_{1,b}$ (which could be for instance ribosomes or RNA polymerases or some of their intermediates) with respect to unbound active individuals $B_{1,u}$. The main interest of this definition is that it is independent of the type of action of the antibiotic and can be used to compare the efficiency of different antibiotics. It captures the inhibition of protein synthesis by the drug, which is also correlated to the cidality of this drug \cite{svetlov_kinetics_2017}. 
Naturally, other choices could be  possible for a proxy of cell risk. Another possible choice would be to compare the ratio of $B_{1,u}$ in the presence and in the absence of antibiotics. A disadvantage of such a definition however is that it would require a choice of reference point for what low risk means and a characterization of that state. 

Instead, with the proposed definition above, in terms of the fraction of active ribosomes, we have a direct link with a quantity that controls the production of proteins \cite{chaturvedi_protein_2016,kohanski_mistranslation_2008}. 
Note that in the previous section, we analyzed Kanamycin, which is known to be bacteriocidal with the same framework we used for bacteriostatic antibiotics. This supports the idea of a proxy of cell risk applicable across various antibiotics types, at least in the early regime following the application of the drug where the reduction of growth is the main effect. 

\begin{figure*}[h!]
    \centering
    \includegraphics[width=\textwidth]{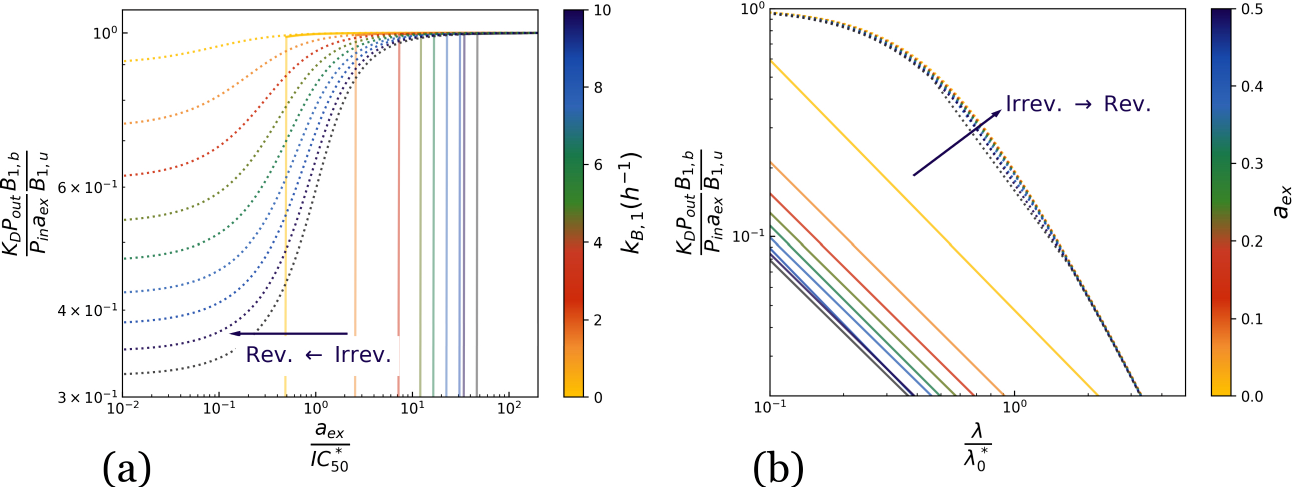}
    \caption{Normalized risk versus antibiotic concentration. (a) Risk faced by the system in the presence of a toxic agent. We compare the reversible case (dotted lines) and the irreversible case (full lines). (b) Rescaled risk depending on the growth rate. We compare the reversible case (dotted lines) and the irreversible case (full lines). We observe a complete collapse of the curves in the reversible limit. The risk is rescaled by $\frac{K_D P_{out}}{P_{in}a_{ex}}$.}
    \label{fig:risk}
\end{figure*}

We show the prediction for this proxy of risk that follows from our model for the reversible and irreversible regimes in Fig.\ref{fig:risk}. 

This proxy of risk has a simple expression
\begin{equation}
\begin{split}
    \frac{B_{1,b}}{B_{1,u}} \simeq \frac{k_{B1}}{\lambda}-1 ,
\end{split}
\label{eq:risk_simple}
\end{equation}
when ribosomes have a long lifetime, fast assembly and fast activation. Note that in Fig.\ref{fig:risk}, the concentration of toxic agent is rescaled by a typical concentration inspired from \cite{greulich_growth-dependent_2015}, $IC_{50}^*=\frac{\sqrt{K_D P_{out} k_{B,1}}}{P_{in}}$.

As expected, the risk is increasing with the fraction of toxic agent while it is decreasing with $\lambda_0$. The risk increases rapidly close to $IC_{50}^*$, with a discontinuity at a given fraction $a_{ex,lim}$ in the irreversible case. This fraction can be understood as a limit concentration above which the system is significantly endangered. In Fig.\ref{fig:risk}, we rescale the risk by $P_{in}a_{ex}/(K_DP_{out})$ to obtain a collapse of the experimental data in the reversible limit. Indeed for $\lambda/\lambda_0^* \to 0$, the risk is equivalent to $P_{in}a_{ex}/(K_DP_{out})$ in the reversible limit which follows from Eq.\ref{eq:risk_simple}.

\section*{Half-inhibitory concentration}

\begin{figure*}
    \centering
    \includegraphics[width=\textwidth]{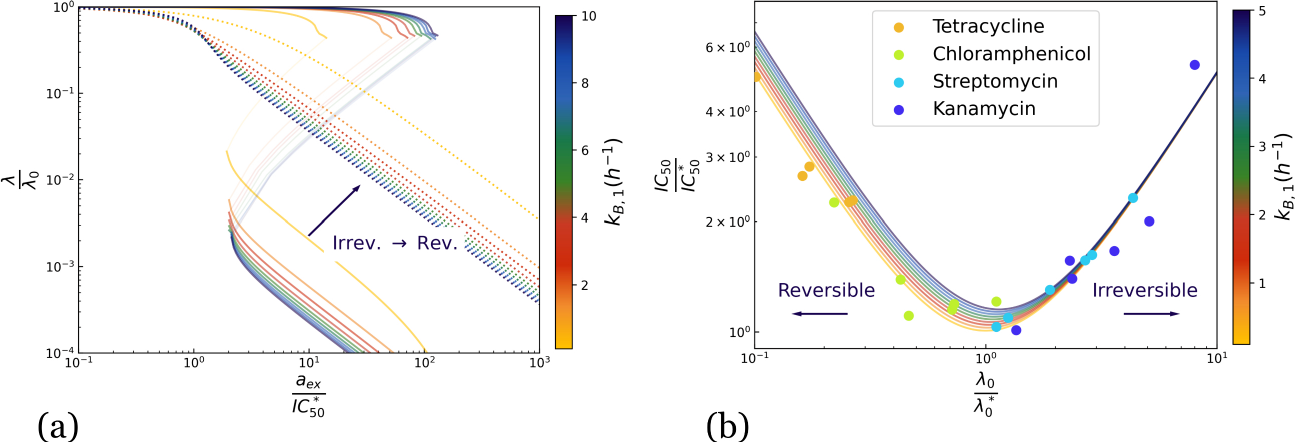}   
    \caption{
    (a) Normalized growth rate versus the normalized antibiotic concentration. In dotted lines we represent the reversible regime $k_{off},P_{out} \geq k_{on}, P_{in}$, in full lines the irreversible regime $k_{off},P_{out} \ll k_{on}, P_{in}$. For the irreversible case (full lines), we observe two branches that represent the coexistence of two values of the growth rate, a "large" growth rate and a "near-zero" growth rate. A discontinuity appears when the system jumps from one branch to another. The colors of the curves correspond to different choices of rate constant $k_{B1}$ as shown on the scale on the right. $k_{B,1}$ essentially sets the basal growth rate $\lambda_0$ (see Supplementary material \cite{SM}) and may vary from one cell to another in a population \cite{kiviet_stochasticity_2014}. \newline
    (b) Half-inhibition concentration $IC_{50}$ as function of the normalized pre-exposure growth rate in the case of no intermediate steps $m=0$. Symbols represent experimental data points extracted from Ref. \cite{greulich_growth-dependent_2015}, which correspond to various antibiotics as shown in the legend.
    }
    \label{fig:ic50_m0}
\end{figure*}

Similarly to what was done in the first model presented, one can study the half-inhibitory concentration $IC_{50}$ with a model based on autocatalytic cycles, assuming they contain an arbitrary number of steps $N$ in the limit of long lifetime and fast assembly.  If we can lump all intermediates into just one ($N=3$), we obtain 

\begin{equation}
    \frac{IC_{50}}{IC_{50}^*}=\frac{1}{2}\left(\left(\frac{\lambda_0^*}{\lambda_0}+2K_D\frac{\lambda_0}{\lambda_0^*}\right)\left(1+\frac{\lambda_0}{2k_{off}}\right) + \frac{\lambda_0}{\lambda_0^*}\right),
\end{equation}
where we have rescaled the half-inhibition concentration by a typical concentration $IC_{50}^*=\sqrt{K_D P_{out} k_{B,1}}/P_{in}$ and the basal growth rate by $\lambda_0^*$. Note that this expression does not depend only on the ratio $\lambda_0/\lambda_0^*$ but also on $\lambda_0$ (itself defined by the parameters of the system). The rescaled half-inhibition concentration as a function of the rescaled basal growth rate in this limit is the convex function shown in Fig.\ref{fig:ic50_m0}b. Remarkably, this function allows to collapse the measurements of many types of antibiotics in a way which is similar with what was done in Ref \cite{greulich_growth-dependent_2015} (for this comparison the same experimental data has been used).
Note also that in the limit of long lifetime, fast binding, fast assembly, and with $k_{off}\gg \lambda_0$, we recover the universal growth dependent susceptibility curve of Eq. \ref{Greulich_universal}.

For an arbitrary number of steps, we also recover in Fig.\ref{fig:ic50_m0}b the two regimes of antibiotics binding mentioned before, namely the reversible regime where the half-inhibitory concentration decreases with $\lambda_0$ and the irreversible regime where it increases with $\lambda_0$. Adding intermediate steps in the autocatalytic cycle shifts the minimum of the parabola towards lower $\lambda_0$ and reduces $IC_{50}$ and thus makes it easier to inhibit growth in the cycle. It also introduces a stronger dependence of $IC_{50}$ on the rate constants $k_{1,B}$ in the reversible regime as compared to the irreversible regime. This reflects that intermediate steps have a stronger impact in reversible pathways as compared to irreversible ones. 

\section*{Extensions of the model}

Other phenomena can be treated with our framework, while we can not be exhaustive, we study two specific extensions: in the first one we consider the combined action of two antibiotics that target two coupled autocatalytic cycles and in the other one, we consider the possibility that the product of one cycle inhibits that cycle.

\subsection*{Effect of two antibiotics targeting two coupled autocatalytic cycles}
\label{ap:cumulatuve_eff}

\begin{figure}[h!]
    \centering
    \includegraphics[width=\linewidth]{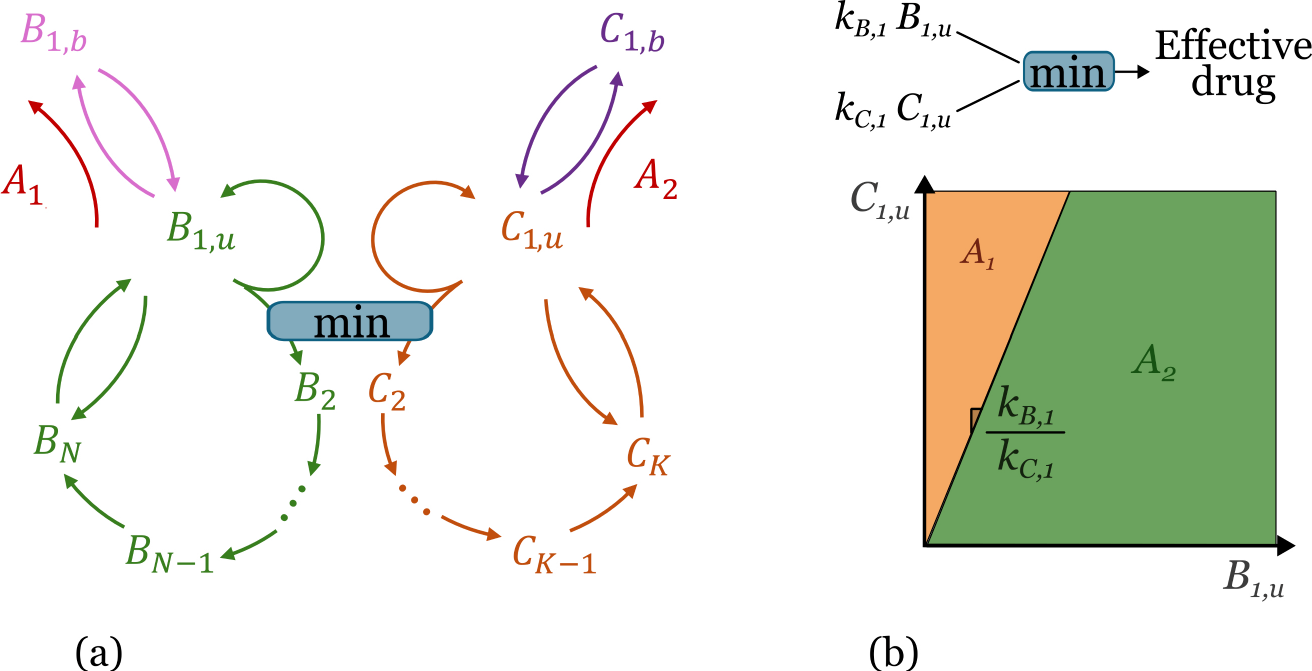}
    \caption{(a) Two antibiotics $A_1$ and $A_2$ targeting two different but coupled autocatalytic cycles (coupled through a $\min$ function represented with the blue box). (b) Predominance diagram of the two drugs, when $B_{1,u}$ (resp. $C_{1,u}$) gets small, the associated cycle is limiting and the effective antibiotic is the one targeting this cycle.}
    \label{fig:network_cumulative}
\end{figure}

The combined action of two antibiotics targeting simultaneously the same ribosomes (and thus the same autocatalytic cycle) has been studied theoretically in \cite{kavcic_minimal_2021} and experimentally validated in \cite{kavcic_mechanisms_2020}. Interestingly, the authors found different regimes of drug interactions such as synergy (the combined effect is stronger) and antagonism (the combined effect is weaker). Another possibility is that the combined effects of the two drugs can be less than that expected based on the individual effect of the drugs, in which case one speaks of suppressive effects \cite{bollenbach_nonoptimal_2009}.
Inspired by these works, we now apply our framework to study the effect of two antibiotics $A_1$ and $A_2$ targeting molecules belonging to separate but coupled autocatalytic cycles as sketched in Fig.\ref{fig:network_cumulative}.

\begin{figure}[h!]
    \centering
    \includegraphics[width=\linewidth]{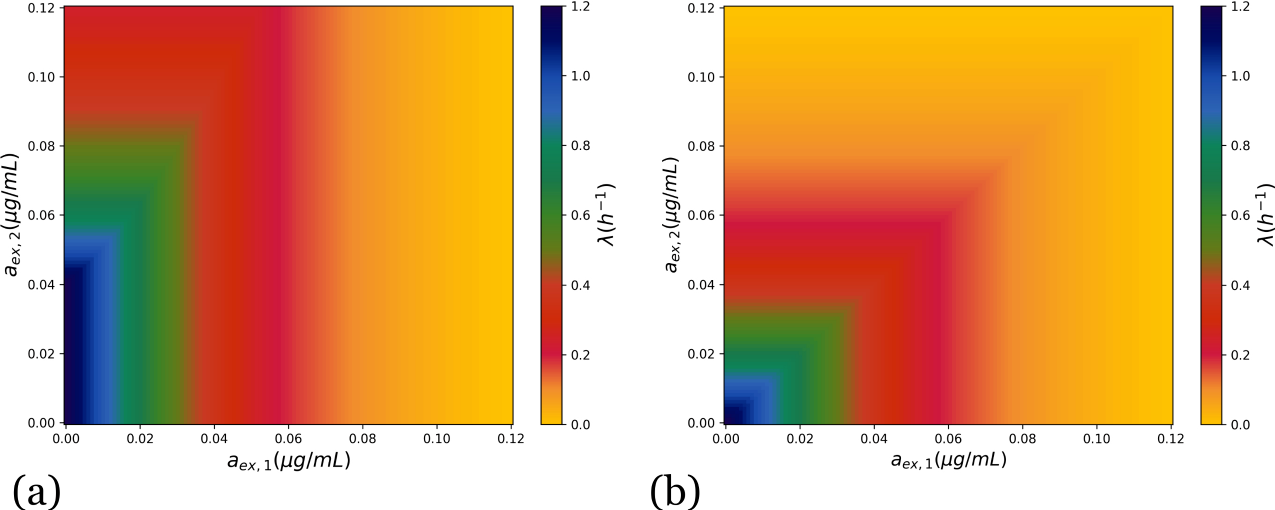}
    \caption{Dose response surface of two antibiotics targeting two coupled autocatalytic cycles. On {(a)}, the drug 1 is in the \textit{reversible regime} with parameters: $P_{in (B)}=40 mL.\mu g .h^{-1}$, $P_{out (B)}=30 h^{-1}$, $k_{B,1}=2 h^{-1}$, $\tau_{life(B)}=10^2 h$, while drug 2 is in the \textit{irreversible regime} with parameters:
    $P_{in (C)}=0,7 mL.\mu g .h^{-1}$, $P_{out (C)}=2 h^{-1}$, $k_{C,1}=1 h^{-1}$, $\tau_{life(C)}=10^2 h$. On {(b)} instead, both drugs are in the \textit{reversible} regime with parameters $P_{in}=40 mL.\mu g .h^{-1}$, $P_{out}=30 h^{-1}$, $k_{B,1}=k_{C,1}=1 h^{-1}$, $\tau_{life(B)}=\tau_{life(C)}=10^2 h$.}
    \label{fig:cumulative_inhibition}
\end{figure}

Interactions effects between the two drugs can be quantified by the dose response surface, which represents the growth rate as function of both drug concentrations as shown in Fig.\ref{fig:cumulative_inhibition}. In the case we consider, the two autocatalytic cycles are coupled with a minimum function introduced previously. As a result, no synergy of the antibiotics is possible because the system behaves as if only one antibiotic was active for a given set of $(a_{ex,1}, a_{ex,2})$. Thus, we obtain an \textit{antagonistic} \cite{kavcic_minimal_2021} interaction between the two drugs because the effect of the drug acting on the limiting cycle is the only one decreasing the growth rate (the second drug has no effect on the growth rate as long as the targeted cycle is not limiting). On Fig.\ref{fig:cumulative_inhibition}, we also see that the transition from one drug to the other depends on the regime of action of each drug. On Fig.\ref{fig:cumulative_inhibition}a, drug $1$ operates in the reversible regime whereas drug $2$ operates in the irreversible regime. If we fix $a_{ex,1}$ (resp. $a_{ex,2}$) and increase $a_{ex,2}$ (resp. $a_{ex,1}$), the growth rate will be constant until $a_{ex,2}$ (resp. $a_{ex,1}$) becomes large enough for drug $2$ (resp. drug $1$) to be the inhibiting drug, and it will start decreasing. We also observe that at  fixed $a_{ex,1}$, the growth rate decreases more slowly with $a_{ex,2}$ than it does with $a_{ex,1}$ at fixed $a_{ex,2}$. 

When the parameters are the same for the two drugs, the transition between the domains of predominance of each antibiotic is along the diagonal $a_{ex,1} = a_{ex,2}$. On Fig.\ref{fig:cumulative_inhibition}a, this transition is shifted upward as compared to Fig.\ref{fig:cumulative_inhibition}b where the two drugs are both reversible with the same parameters. Indeed on Fig.\ref{fig:cumulative_inhibition}a, as drug $1$ produces a stronger inhibition than drug $2$ at the same concentration, drug $1$ is effective even in regions where $a_{ex,1}<a_{ex,2}$.

As shown in calculations detailed in Supplementary material \cite{SM} Section C, the minimum law sets a transition between the effect of one antibiotic and the effect of the other one, depending on which cycle becomes limiting (when $a_{ex,1} \gg a_{ex,2}$, the cycle inhibited by $a_{ex,2}$ becomes limiting and vice versa). Actually, the transition does not occur at $a_{ex,1} = a_{ex,2}$ but when both terms of the minimum function shown in the Fig. \ref{fig:network_cumulative} are identical.

\subsection*{Closed compartment and inhibiting waste}
In this section we show that our model can describe other systems than cycles inhibited by antibiotics. In particular we consider the network represented on Fig.\ref{fig:risk_waste}, in which a waste $W$ is produced at a rate $k_w$ (see Supplementary material \cite{SM} Section C). This waste then inhibits autocatalysts by binding to them in a similar fashion than antibiotics in the previous sections. We consider only a closed compartment $P_{in}=P_{out}=0$, meaning that waste only comes from the cycle itself and never leaves the compartment, then the risk is

\begin{equation}
    \frac{B_{1,b}}{B_{1,u}}=\frac{k_{on}k_wQ(\lambda)}{\left(\lambda+k_{off}\right)\left(\lambda+k_{on}Q(\lambda)\frac{\lambda}{\lambda+k_{off}}\right)},
\end{equation}

\noindent where $Q(\lambda)$ was defined in Eq.\ref{eq:Qlambda}. Interestingly, there are regimes where the risk is an increasing function of the growth rate $\lambda$ as shown on Fig.\ref{fig:risk_waste}. This regime corresponds to an accumulation of bound individuals when the growth rate is increasing, which are not diluted fast enough.

\begin{figure}[h]
    \centering
    \includegraphics[width=\textwidth]{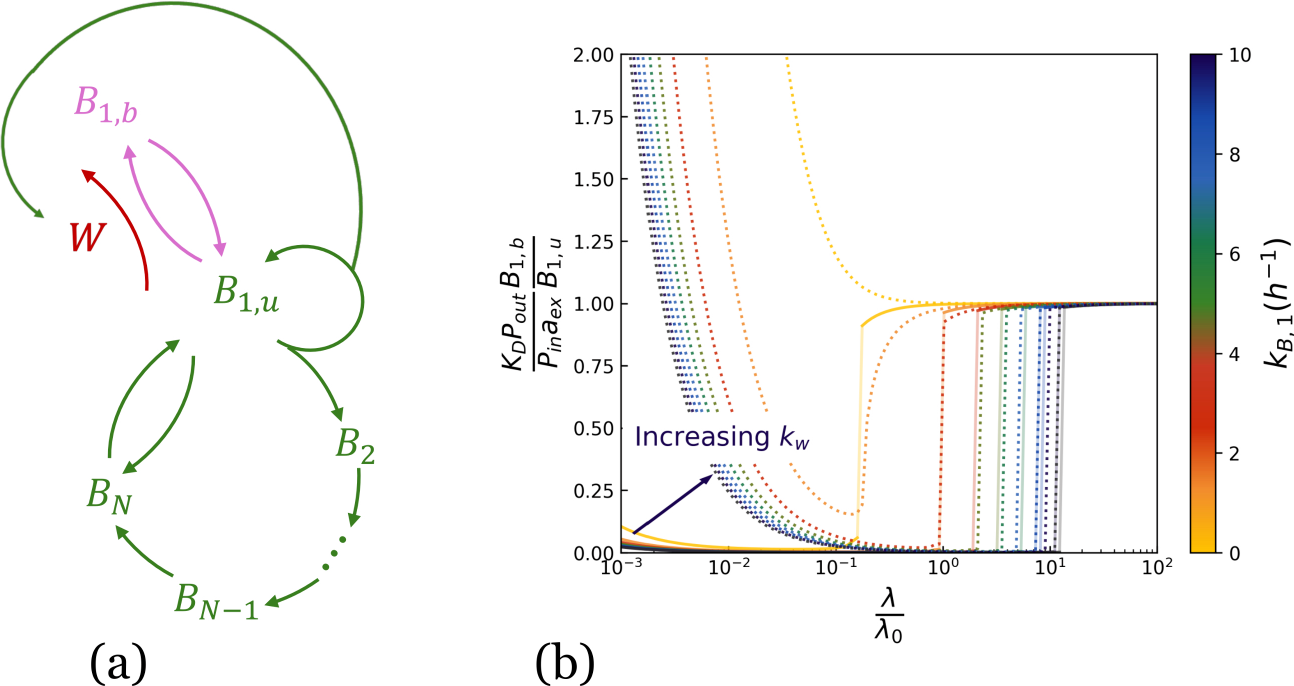}
    \caption{(a) Network where self inhibiting waste is produced. (b) Risk related to growth in a regime where risk can be increasing with $\lambda$. The full lines corresponds to a higher value of $k_w$ ($k_w=0.01h^{-1}$) compared to the dotted lines ($k_w=1h^{-1}$).}
    \label{fig:risk_waste}
\end{figure}

Some biological processes where the bacteria produces a self-inhibitory compound may be modeled in such a way, for instance the production of ribosome modulation factor \cite{wada_ribosome_1995} or inhibition due to by-products in yeast \cite{zentou_new_2021}.

\section*{Conclusion}
We constructed a minimal biophysical model for the inhibition of bacterial growth by antibiotics based on a model of cell metabolism in terms of coupled autocatalytic cycles, that can contain an arbitrary number of steps.
Unlike what was done in Ref. \cite{greulich_growth-dependent_2015}, our approach does not assume growth laws, instead they are derived from the model. The model describes well the effects of a large panel of antibiotics targeting key autocatalytic cycles in E.Coli. 
We have found that the two regimes previously identified for ribosome-targeting antibiotics in \cite{greulich_growth-dependent_2015}, namely the reversible (strong outflux of inhibitors) and irreversible (small outflux of inhibitors) regimes, should in fact be expected generically for any inhibitors targeting an autocatalytic cycle. Further, we found a region of growth rate heterogeneity in a certain range of parameters, which has been reported experimentally. We were also able to describe the antagonistic effect of two drugs targeting different autocatalytic cycles, and self-inhibition from a toxic by-product.

In the future, we would like to expand our approach towards bacteriocidal antibiotics, which are typically used in conjunction with bacteriostatic antibiotics in a time-dependent manner \cite{marrec_resist_2020}. 
The observation that our model successfully describe the effect of Kanamycin although this antibiotics is classified as bacteriocidal suggests that bactoriocidal antibiotics may also lower the growth rates similarly as bacteriostatic antibiotics before they kill the cell. This would imply that the two classes of antibiotics may be more similar than previously thought and that the proxy of risk which we have introduced could be a relevant measure to classify antibiotics irrespective of the class to which they belong.

We are planning to also study experiments that show significant cell-to-cell heterogeneity in antibiotic susceptibility \cite{irwin_evidence_2010,le_quellec_measuring_2024}. Modeling these experiments will require a stochastic version of the present model which is needed for describing the growth and death of individual cells and the stochastic effects due to population size.
In this respect, it is encouraging to see that our model predicts growth rate heterogeneity even in the absence of noise, but 
single-cell experiments are needed to analyze this growth rate heterogeneity more precisely and to relate the single-cell behavior to population susceptibility.

We have also explored the question of drug interactions inspired by Ref. \cite{kavcic_minimal_2021}, which was also built on \cite{greulich_growth-dependent_2015} and therefore also used  \textit{phenomenological growth laws} as assumptions. Since our model relied instead on a more detailed modeling of translation and transcription using autocatalytic cycles, we were able to study unexplored mechanism of action of antibiotics, such as the case where the two antibiotics target two coupled cycles, which could be ribosomal and m-RNA for instance. In this case, we predict antagonist interactions among the two drugs and a dose response surface which depends whether the two drugs are in the same binding regime or in different ones.

Finally, let us also point out that our approach based on autocatalytic cycles is rather general and could be applied beyond cellular biology to other fields, such as ecology \cite{veldhuis_ecological_2018} or economy, where individuals rather than molecules are able to create more of themselves thanks to autocatalytic cycles but can also be inhibited by  toxic agents, either present in their environment or created by themselves as a result of their own growth, a case we also considered in the last extension of this framework. 

\section*{Acknowledgements}
We acknowledge inspiring  discussions with C. Baroud and E. Maikranz, and a critical reading  by L. Dinis.

\appendix

\section{Definition of the model and derivation of the growth laws} \label{ap:derivation_simple_case}

The law of the minimum used here to model cell metabolism actually emerges from works in economy and are related to the Leontief's production function. This law concerns the formation of a product $P$, for which $n_1$ units of $R_1$, $n_2$ units of $R_2$, ... up to $n_N$ units of $R_N$ are assembled. The production rate of $P$ is limited by the smaller value of $R_j/n_j$, that is the number of sets of resource $j$ required to produce $P$. It also depends on the minimum time to produce one unit of $P$ $\tau_P$, and the minimum time to use resource $R_j$ in order to produce one $P$ $\tau_i$. If resources are not fully allocated to the production of one product $P$, and several products $P_i$ are assembled in parallel, one resource may be used by different production chains simultaneously. In this case a fraction $\alpha_{i,j}$ of total available resources $R_j$ must be used to produce $P_i$, so that

\begin{equation}
    \frac{dP_i}{dt} = \frac{1}{\tau_P}\min \left( \alpha_{i,1}\frac{\tau_P}{\tau_1}\frac{R_1}{n_1}, \alpha_{i,2}\frac{\tau_P}{\tau_2}\frac{R_2}{n_2}, ..., \alpha_{i,N}\frac{\tau_P}{\tau_N}\frac{R_N}{n_N} \right).
\end{equation}

where $\alpha_{i,j}\tau_P/\tau_j n_j$ represents the maximal number of copies of the product that you can produce simultaneously from one unit of resource $i$. In our model, $1/k_{B,1}$ (resp. $1/k_{C,1}$) is the minimal time required to use $B_{1,u}$ (resp. $C_{1,u}$) in order to increase either $B_2$ or $C_2$.
 
Within Leontief's approach \cite{dobos_dynamic_2005}, or Liebig's model in ecology \cite{oneill_multiple_1989}, the rates of reactions involving two complementary resources are set by the limiting quantity among the two using a minimum function as shown in Fig. 1a. 
We use numbers rather than concentrations in order to use the law of the minimum of Leontief's formalism. In particular, for a number $N$ of individuals in a volume $\Omega$, the concentration is $c=N/\Omega$. We get: 
\begin{equation}
	\frac{d c}{d t} = \frac{1}{\Omega} \frac{dN}{dt} - c \frac{1}{\Omega}\frac{d\Omega}{dt},
\end{equation}
\noindent where the second term is the so called \textit{"dilution term"} $-\lambda c$, with the growth rate $\lambda =  ({1}/{\Omega}) ({d\Omega}/{dt})$. Therefore, if we assume steady states for the concentration $c$, we get:
\begin{equation}
	\frac{dN}{dt} = \lambda N.
\end{equation}

\subsection{Simplified model}
Here, we consider the network shown on Fig.1, in which at most three intermediates are present for ribosomes or RNA precursors. We will later relax this assumption and consider an arbitrary number of intermediates:

\begin{equation}
\label{eq:starting_eqs}
\begin{split}
    \frac{dB_{1,u}}{dt} &= k_{B3}B_3 - k_{B4} B_{1,u} - \hat{k}_{on} \frac{A}{\Omega} B_{1,u} + k_{off} B_{1,b} - \frac{B_{1,u}}{\tau_{life}}
    \\ \frac{dB_{1,b}}{dt} &= \hat{k}_{on} \frac{A}{\Omega}  B_{1,u} - k_{off} B_{1,b} - \frac{B_{1,b}}{\tau_{life}}
    \\ \frac{dB_2}{dt} &= \min(k_{B1}B_{1,u},k_{C1}C_1) - k_{B2} B_2 - \frac{B_2}{\tau_{life}}
    \\ \frac{dB_3}{dt} &= k_{B2}B_2 - k_{B3}B_3 + k_{B4}B_{1,u} - \frac{B_3}{\tau_{life}}
    \\ \frac{dC_1}{dt} &= k_{C3}C_3- k_{C4} C_1 - \frac{C_1}{\tau_{life(C)}}
    \\ \frac{dC_2}{dt} &= \min(k_{B1}B_{1,u},k_{C1}C_1) -  k_{C2} C_2 - \frac{C_2}{\tau_{life(C)}}
    \\ \frac{dC_3}{dt} &= k_{C2}C_2 - k_{C3}C_3 + k_{C4}C_1 - \frac{C_3}{\tau_{life(C)}}
    \\ \frac{dA}{dt} &= \hat{P}_{in} a_{ex} \Omega - P_{out} A - \hat{k}_{on} \frac{A}{\Omega} B_{1,u} + k_{off} B_{1,b},
\end{split}
\end{equation}
where $k_i$ and $\hat{k}_i$ are rate constants. 
We now introduce the ribosome concentration $\rho$ such that 
$\Omega =  B_{tot} / \rho$. 
This ribosome concentration is assumed to be a constant \cite{neurohr_relevance_2020}, which does not depend on the antibiotic concentration.
Thus,
we can absorb the factor $\rho$ into $k_{on}$ using $k_{on}= \hat{k}_{on} \rho$ and similarly with $P_{in}= \hat{P}_{in}/\rho$. When the species $B$ is limiting, the minimum function can be simplified, the equations for $C_1$, $C_2$ and $C_3$ may be discarded and we get a simpler system:

\begin{equation}
\begin{split}
    \frac{dB_{1,u}}{dt} &= k_{B3}B_3 - k_{B4} B_{1,u} - k_{on} \frac{A}{B_{tot}} B_{1,u} + k_{off} B_{1,b} - \frac{B_{1,u}}{\tau_{life}}
    \\ \frac{dB_{1,b}}{dt} &= k_{on} \frac{A}{B_{tot}}  B_{1,u} - k_{off} B_{1,b} - \frac{B_{1,b}}{\tau_{life}}
    \\ \frac{dB_2}{dt} &= k_{B1}B_{1,u} - k_{B2} B_2
    \\ \frac{dB_3}{dt} &= k_{B2}B_2 - k_{B3}B_3 + k_{B4}B_{1,u} - \frac{B_3}{\tau_{life}}
    \\ \frac{dA}{dt} &= P_{in} B_{tot} a_{ex} - P_{out} A - k_{on} \frac{A}{B_{tot}} B_{1,u} + k_{off} B_{1,b}.
\end{split}
\end{equation}

\noindent Let now assume that this system has a largest eigenvalue $\lambda$, which describes exponential growth. Since we are interested in a regime of balanced growth, this factor $\lambda$ also represents the dilution rate that follows from the growth of the cell volume. Let us then also assume that the life time of the ribosome precursors $\tau_{life}$ is long with respect to $1/\lambda$. In that case we obtain the system

\begin{equation}
\begin{split}
    \left(\lambda+k_{on}\frac{A}{B_{tot}}+k_{B4}\right) B_{1,u} &= k_{B3}B_3 + k_{off} B_{1,b}
    \\ \left(\lambda  + k_{off}\right)B_{1,b} &= k_{on} \frac{A}{B_{tot}} B_{1,u}
    \\ \left(\lambda + k_{B2}\right)B_2 &= k_{B1}B_{1,u}
    \\ \left(\lambda+k_{B3}\right)B_3 &= k_{B2}B_2 + k_{B4}B_{1,u}
    \\ \left(\lambda + P_{out} + k_{on} \frac{B_{1,u}}{B_{tot}}\right)A &= P_{in} B_{tot} a_{ex} + k_{off} B_{1,b}.
\end{split}
\label{eq:system1}
\end{equation}

\noindent We now normalize all quantities with respect to the total amount of mature $B$ molecules, $B_{tot} = B_{1,u} + B_{1,b} + B_3$. 
We find by summing equations 1, 2 and 4 of the previous system:

\begin{equation}
\begin{split}
    \lambda \left(B_{1,u}+B_{1,b}+B_3\right)  =  k_{B2}B_2, 
\end{split}
\label{eq:btot_combin}
\end{equation}
which is equivalent to $\lambda B_{tot}=k_{B2}B_2$.

\noindent From the other equations, we have (third equation of Eq.\ref{eq:system1} and definition of $B_{tot}$):

\begin{equation}
\begin{split}
    (\lambda+k_{B2})B_2 &=k_{B1}B_{1,u}
    \\ B_{1,b} = B_{tot}\hspace{-0.1cm}-\hspace{-0.1cm}B_{1,u}\hspace{-0.1cm}-\hspace{-0.1cm}B_3 &= B_{tot}\hspace{-0.1cm}-\hspace{-0.1cm}B_{1,u}\hspace{-0.1cm}-\hspace{-0.1cm}\frac{k_{B2}B_2+k_{B4}B_{1,u}}{\lambda+k_{B3}}.
\end{split}
\label{eq:system2}
\end{equation}

\noindent From this, we recover the equivalent of the first growth law for ribosomes (combining Eq.\ref{eq:btot_combin} and the first of Eq.\ref{eq:system2}):
\begin{equation}
    \frac{B_{1,u}}{B_{tot}}=\frac{\lambda}{k_{B1}}  \left(1+\frac{\lambda}{k_{B2}}\right).
\label{eq:b1u1}
\end{equation}
To simplify the calculations, we  introduce the notation $Q(\lambda) := B_{1,u}/B_{tot}$ in the following.

\noindent The other equations give
\begin{equation}
\begin{split}
    \frac{B_2}{B_{tot}} &= \frac{\lambda}{k_{B2}},
    \\ \frac{B_{1,b}}{B_{tot}} &= 1\hspace{-0.1cm}-\hspace{-0.1cm}\frac{\lambda}{k_{B1}}(1+\frac{\lambda}{k_{B2}})\hspace{-0.1cm}-\hspace{-0.1cm}\frac{\lambda}{\lambda+k_{B3}}\hspace{-0.1cm}-\hspace{-0.1cm}\frac{k_{B4}\lambda (1+\frac{\lambda}{k_{B2}})}{k_{B1}(\lambda+k_{B3})}.
\end{split}
\label{eq:b1b1}
\end{equation}
Using the second equation of Eq. \ref{eq:system1}, we can write $B_{1,b}$ in another way:

\begin{equation}
\begin{split}
    B_{1,b}&=\frac{k_{on}AB_{1,u}}{B_{tot}(\lambda+k_{off})} = \frac{k_{on} A Q(\lambda)}{\lambda + k_{off}},
\end{split}
\end{equation}

\noindent and compute explicitly the abundance of antibiotics from the last equation of Eq.\ref{eq:system1}:

\begin{equation}
    A=\frac{P_{in} B_{tot} a_{ex}}{\lambda + P_{out}+ \frac{k_{on}\lambda Q(\lambda)}{\lambda+k_{off}}}.
\end{equation}

\noindent Now we can eliminate $A$ from the previous two equations, which leads to a new expression for $B_{1,b}$:

\begin{equation}
    B_{1,b}=\frac{ P_{in}a_{ex} B_{tot} k_{on} Q(\lambda)}{\left(\lambda+k_{off}\right)\left(\lambda + P_{out} \right) + k_{on} Q(\lambda) \lambda}.
\label{eq:b1b2}
\end{equation}

\subsection{Inhibitor-free growth rate}
\noindent Without toxic agent ($a_{ex}=0$), we obtain from Eq.\ref{eq:b1b2} $B_{1,b}=0$, which implies using Eq.\ref{eq:b1b1} the equation
\begin{equation}
    k_{B1}k_{B3}=\lambda_0 \left(\lambda_0+k_{B3}+k_{B4}\right)\left(1+\frac{\lambda_0}{k_{B2}}\right),
\end{equation}
where, $\lambda_0$ is the value of $\lambda$ in the absence of inhibitor, i.e. the "inhibitor-free" growth rate of the cell. As the concentration of antibiotics increases, the growth rate is modified. In particular, we always have $\lambda \leq \lambda_0$ for bacteriostatic drugs.

\subsection{Second growth law}
\noindent To recover the second growth law, we simply sum Eq.\ref{eq:b1u1} and Eq.\ref{eq:b1b1}:
\begin{equation}
\label{eq:ratio}
\begin{split}
    \frac{B_{1,u}+B_{1,b}}{B_{tot}}&=\frac{k_{B3} - k_{B4} Q(\lambda)}{\lambda+k_{B3}}.
\end{split}
\end{equation}

\noindent In the limit of fast assembly ($k_{B2},k_{B3}\gg\lambda$), 
we find $Q(\lambda) \simeq \lambda/k_{B1}$ and 
\begin{equation}
    \frac{B_{1,u}+B_{1,b}}{B_{tot}}=1-\frac{\lambda}{k_{B3}} \left( 1 + \frac{k_{B4}}{k_{B1}} \right),
\end{equation}
which assuming in addition fast activation ($k_{B4}\ll k_{B1}$) further simplifies in
\begin{equation}
    \frac{B_{1,u}+B_{1,b}}{B_{tot}}=1-\frac{\lambda}{k_{B3}}.
\end{equation}
Note that this model always predicts a negative correlation between the growth rate and the ratio $(B_{1,u}+B_{1,b})/B_{tot}$ if the growth rate is high enough from  Eq. \ref{eq:ratio} because $Q(\lambda)$ is a quadratic function of $\lambda$. In the limit of fast assembly ($k_{B2},k_{B3}\gg\lambda$), this correlation takes the form of a linear dependence in $\lambda$ in agreement with \cite{scott_interdependence_2010}.

\subsection{Self-consistent equation for the growth rate}
\label{sec: self-consist}
Without any assumptions on the rates, equating the two equations for $B_{1,b}$ (Eq.\ref{eq:b1b1} and Eq.\ref{eq:b1b2}) yields the self-consistent equation for the growth rate:
\begin{equation} 
\label{eq:self-consist}
    \frac{P_{in}a_{ex} k_{on} Q(\lambda)}{ ( \lambda + P_{out} ) ( \lambda + k_{off} ) + k_{on} \lambda Q(\lambda) } = \frac{k_{B3} - (k_{B3}+k_{B4}+\lambda) Q(\lambda)}{\lambda+k_{B3}}.
\end{equation}
In order to obtain a more manageable expression, we now assume: $k_{B3} \gg k_{B4}$ and ($k_{B2},k_{B3}\gg\lambda_0$), which lead to 
$\lambda_0 \simeq k_{B1}$ and
$Q(\lambda) \simeq \lambda/\lambda_0$. 
These approximations are expected to hold for ribosomes which can be described by long lifetimes, fast assembly and fast activation rates.
Since $\lambda < \lambda_0$, this approximation also implies 
($k_{B2},k_{B3}\gg\lambda$), and therefore Eq. \ref{eq:self-consist} takes the simpler form of a cubic equation for $\lambda$:
\begin{equation}
\label{eq:self-consist-simple}
    P_{in}a_{ex} k_{on} \frac{\lambda}{\lambda_0} \hspace{-0.1cm}= \hspace{-0.1cm}\left( 1 - \frac{\lambda}{\lambda_0} \right)\hspace{-0.15cm} \left[  ( \lambda + k_{off} ) (\lambda+ P_{out} ) \hspace{-0.1cm}+\hspace{-0.1cm} k_{on} \frac{\lambda^2}{\lambda_0} \right].
\end{equation}


\subsubsection{Reversible limit}

Let us now introduce a typical growth rate $\lambda_0^*=2\sqrt{P_{out} K_D \lambda_0}$ where $K_D=k_{off}/k_{on}$.
In the reversible limit defined by  $\lambda \ll \lambda_0^*$, one also has $P_{out},k_{off}\gg \lambda$ and thus Eq. \ref{eq:self-consist-simple} leads to
\begin{equation}
    \frac{\lambda}{\lambda_0} \left(K_D P_{out} + P_{in}a_{ex} \right)  = K_D P_{out},
\end{equation}
and therefore:
\begin{equation}
    \lambda=\frac{\lambda_0}{1+\frac{P_{in}a_{ex}}{K_D P_{out}}},
\end{equation}
which is the result obtained in \cite{greulich_growth-dependent_2015} for the reversible case.

\subsubsection{Irreversible limit}

In the irreversible limit instead,  $\lambda \gg \lambda_0^*$. This implies   
$P_{out},k_{off} \ll \lambda$ and $k_{on} \gg \lambda_0$, and 
Eq. \ref{eq:self-consist-simple} leads to:
\begin{equation}
    \left(\frac{\lambda}{\lambda_0}\right)^2-\left(\frac{\lambda}{\lambda_0}\right)+\frac{P_{in}a_{ex}}{\lambda_0}=0.
\end{equation}
In this case:
\begin{equation}
    \lambda=\frac{\lambda_0}{2}\left(1+\sqrt{1-\frac{4P_{in}a_{ex}}{\lambda_0}}\right),
\end{equation}
also in agreement with \cite{greulich_growth-dependent_2015}.

\subsection{General case: arbitrary number of intermediate construction steps}\label{subsec:general_arbitrary_number}
\label{ap:derivation_arbitrary_number}
For some processes (such as the autocatalytic cycle of RNA polymerase \cite{roy_unifying_2021}), some intermediate steps can be be significant to form mature autocatalysts $B_1$ as sketched on Fig 1a of the main text. As an example, to form RNA-polymerase, mRNA have to be translated to resting protein subunits, that have to be activated and then assembled to form resting RNA-polymerase ($B_{N-1}$ in Fig. 1a, with $N=5$ in this example). Examples from ecology, or economy could involve slow assembly steps affecting the growth rate. Typically, if one sub-unit of the system is produced slowly we expect the system to be limited by this step, whereas fast assembly steps should not influence the growth rate. Here, we extend the previous model to include an arbitrary number of intermediate steps. Below, we do this for the first cycle only, assuming $B$ is limiting as done previously.

\noindent The rate equations now become:

\begin{equation}
\begin{split}
    \frac{dB_{1,u}}{dt} &= k_{B,N}B_N - k_{B,N+1} B_{1,u} - k_{on} \frac{A}{B_{tot}} B_{1,u} + k_{off} B_{1,b} - \frac{B_{1,u}}{\tau_{life}}
    \\ \frac{dB_{1,b}}{dt} &= k_{on} \frac{A}{B_{tot}}  B_{1,u} - k_{off} B_{1,b} - \frac{B_{1,b}}{\tau_{life}}
    \\ \frac{dB_2}{dt} &= k_{B,1}B_{1,u} - k_{B,2} B_2
    \\ \vdots
    \\ \frac{dB_{N}}{dt} &= k_{B,N-1}B_{N-1} - k_{B,N}B_{N} + k_{B,N+1}B_{1,u} - \frac{B_{N}}{\tau_{life}}
    \\ \frac{dA}{dt} &= P_{in} B_{tot} a_{ex} - P_{out} A - k_{on} \frac{A}{B_{tot}} B_{1,u} + k_{off} B_{1,u}
\end{split}
\label{eq:general_eq_dyn}
\end{equation}

\noindent With the assumption of exponential growth with a rate $\lambda$ and that of a long life time $1 /\tau_{life} \ll \lambda$, we obtain the system:

\begin{equation}
\begin{split}
    \left(\lambda + k_{B,N+1} + k_{on} \frac{A}{B_{tot}}\right)B_{1,u} &= k_{B,N}B_N + k_{off} B_{
    1,b}
    \\ \left(\lambda  + k_{off}\right)B_{1,b} &= k_{on} \frac{A}{B_{tot}} B_{1,u}
    \\ \left(\lambda + k_{B,2}\right)B_2 &= k_{B,1}B_{1,u}
    \\ \vdots
    \\ \left(\lambda + k_{B,N-1} \right)B_{N-1} &= k_{B,N-2}B_{N-2}
    \\ \left(\lambda +k_{B,N} +  \right)B_{N} &= k_{B,N-1}B_{N-1} + k_{B,N+1}B_{1,u}
    \\ \left(\lambda+P_{out}+k_{on} \frac{B_{1,u}}{B_{tot}} \right)A &= P_{in} B_{tot} a_{ex} + k_{off} B_{1,u},
\end{split}
\end{equation}

\noindent and if we multiply equations $3$ to $N$ together, we find:

\begin{equation}
    B_{1,u}=\frac{\lambda+k_{B,N-1}}{k_{B,1}}\left(\hspace{-0.1cm}1\hspace{-0.1cm}+\hspace{-0.1cm}\frac{\lambda}{k_{B,2}}\hspace{-0.1cm}\right)\hspace{-0.05cm}\times...\times\hspace{-0.1cm}\left(\hspace{-0.1cm}1\hspace{-0.1cm}+\hspace{-0.1cm}\frac{\lambda}{k_{B,N-2}}\hspace{-0.1cm}\right)B_{N-1} .
\end{equation}

\noindent Defining $B_{tot}=B_{1,u}+B_{1,b}+B_N$, we obtain by summing the two first equations and the $N+1$-th:

\begin{equation}
    B_{N-1}=\frac{\lambda}{k_{B,N-1}} B_{tot},
\end{equation}

\noindent and therefore, we get:

\begin{equation}
    \frac{B_{1,u}}{B_{tot}}=\frac{\lambda}{k_{B,1}}\left(1\hspace{-0.1cm}+\hspace{-0.1cm}\frac{\lambda}{k_{B,2}}\right)\times...\times\hspace{-0.1cm}\left(1\hspace{-0.1cm}+\hspace{-0.1cm}\frac{\lambda}{k_{B,N-1}}\right).
\label{eq:general_growthlaw}
\end{equation}

\noindent This is the equivalent of the first growth law \cite{greulich_growth-dependent_2015, roy_unifying_2021, scott_bacterial_2011} in a general case, 
and in that case, $B_{1,u}/B_{tot}$ is a $(N-1)$-th order polynomial in $\lambda$, which we call $Q(\lambda)$.
This polynomial is positive and increasing over $\mathbb{R}^+$.
Now, if all the intermediate processes are sufficiently fast $\forall n \in \{2,...,N-1\}, \lambda \ll k_{B,n}$, we recover the linear law:
\begin{equation}
    B_{1,u}=\frac{\lambda}{k_{B,1}} B_{tot}.
\label{eq:general_linear_growthlaw}
\end{equation}

\noindent  We can also express the concentration of bound individuals $B_{1,b}$:

\begin{equation}
\vspace*{-2.mm}
\begin{split}
    \\\frac{B_{1,b}}{B_{tot}}&=\frac{k_{B,N}-Q(\lambda)(\lambda+k_{B,N}+k_{B,N+1})}{\lambda+k_{B,N}},
\end{split}
\end{equation}

\noindent we further obtain:

\begin{equation}
\begin{split}
    B_{1,u}&=Q(\lambda)B_{tot}
    \\B_{1,b}&=\frac{k_{B,N}-Q(\lambda)(\lambda+k_{B,N}+k_{B,N+1})}{\lambda+k_{B,N}}B_{tot}
    \\B_{1,b}&=\frac{k_{on}A Q(\lambda)}{\lambda+k_{off}}
    \\A&=\frac{P_{in}B_{tot}a_{ex}}{\lambda+P_{out}+k_{on}Q(\lambda)\frac{\lambda}{\lambda+k_{off}}}.
\end{split}
\end{equation}

\noindent The second equation is obtained by writing $B_{1,b}=B_{tot}-B_{1,u}-B_{N}$. Equating the two equations for $B_{1,b}$, we find the general self-consistent equation on the growth rate Eq.\ref{eq:self_consistent_growthrate_general}. In the absence of toxic agent, $a_{ex}=0$, the growth rate $\lambda_0$ is set by:

\begin{equation}
    Q(\lambda_0)\left(\lambda_0+k_{B,N}+k_{B,N+1}\right)=k_{B,N}.
\end{equation}

\noindent As done previously, we can write a second expression for $B_{1,b}$ as proportional to the abundance of toxic agents $A$. Equating the two equations for $B_{1,b}$, we find a general self-consistent equation on the growth rate, which becomes equivalent to Eq. 3 of the main text when there is only one assembly step $(N=3)$:

\begin{equation}
\frac{k_{on}Q(\lambda)P_{in}a_{ex}\left(\lambda+k_{B,N}\right)}{\left(\lambda+k_{off}\right)\left(\lambda+P_{out}+k_{on}Q(\lambda)\frac{\lambda}{\lambda+k_{off}}\right)}=k_{B,N}-Q(\lambda)(\lambda+k_{B,N}+k_{B,N+1}).
\label{eq:self_consistent_growthrate_general}
\end{equation}

\noindent In the absence of toxic agent, $a_{ex}=0$, and the growth rate $\lambda_0$ is set by taking the right side of the equation to be $0$. This is a generalization of the results discussed previously in the simple case.

\subsubsection{Reversible regime}
\noindent In the reversible limit, $k_{off},P_{out}\gg k_{on}, P_{in},...$. In this case Eq.\ref{eq:self_consistent_growthrate_general} becomes:

\begin{equation}
\begin{split}
    &\frac{k_{on}Q(\lambda)P_{in}a_{ex}\left(\lambda+k_{B,N}\right)}{k_{B,N}k_{off}P_{out}}\hspace{-0.1cm}=\hspace{-0.1cm}1\hspace{-0.1cm}-\hspace{-0.1cm}Q(\lambda)(1\hspace{-0.1cm}+\hspace{-0.1cm}\frac{\lambda}{k_{B,N}}\hspace{-0.1cm}+\hspace{-0.1cm}\frac{k_{B,N+1}}{k_{B,N}}),
\end{split}
\label{self_consistent_growthrate_rev}
\end{equation}

\noindent if we further assume fast assembly

\begin{equation}
\begin{split}
    \frac{k_{on}Q(\lambda)P_{in}a_{ex}}{k_{off}P_{out}}=1-Q(\lambda),
\end{split}
\end{equation}

\noindent and therefore:

\begin{equation}
\begin{split}
    Q(\lambda)=\frac{1}{1+\frac{K_DP_{in}}{P_{out}}a_{ex}}.
\end{split}
\end{equation}

\subsubsection{Irreversible regime}
\noindent In the irreversible limit, $k_{off},P_{out}\ll k_{on}, P_{in},...$, the equation becomes:

\begin{equation}
\begin{split}
  &\frac{P_{in}a_{ex}\left(\lambda+k_{B,N}\right)}{\lambda (\lambda + k_{on} Q) }=k_{BN}-Q(\lambda) \left( \lambda+k_{B,N}+k_{B,N+1} \right),
\end{split}
\end{equation}
which simplifies further when assuming fast assembly, i.e. $\lambda \ll k_{BN}$ and $k_{B,N+1} \ll k_{B,N}$. The assumption $k_{B,N+1} \ll k_{B,N}$ is quite natural because the rate $k_{B,N+1}$ corresponds to a transition in which an active ribosome would go back to a precursor form, an unlikely transition when compared to the forward transformation of a precursor to a fully formed ribosome which has the rate $k_{B,N}$.

\begin{equation}
\begin{split}
    Q(\lambda)=1-\frac{P_{in}a_{ex}}{\lambda}.
\end{split}
\end{equation}

\subsubsection{Second growth law}
\label{ap:secondgrowth_ribo}

\noindent We can also recover a linear decreasing law between the growth rate and the ribosome fraction in the general case. With our formalism, we obtain:

\begin{equation}
    \frac{B_{1,u}+B_{1,b}}{B_{tot}}=1-\frac{\lambda}{\lambda+k_{B,N}}-\frac{k_{B,N+1}Q(\lambda)}{\lambda+k_{B,N}}.
\end{equation}

\noindent In the limit of fast assembly, fast activation, we find:

\begin{equation}
    \frac{B_{1,u}+B_{1,b}}{B_{tot}}=1-\frac{\lambda}{k_{B,N}}.
\end{equation}

\noindent Again, we have a linear decreasing correlation.

\subsubsection{Fast assembly}
\noindent If we assume fast assembly  $\forall l \in\{2,...,N\}, k_{B,N+1} \ll k_{B,1},\lambda_0,\lambda\ll k_{B,l}$ we have:

\begin{equation}
    \frac{k_{on}Q(\lambda)P_{in}a_{ex}}{\left(\lambda+k_{off}\right)\left(\lambda+P_{out}+k_{on}Q(\lambda)\frac{\lambda}{\lambda+k_{off}}\right)}=1-Q(\lambda),
\end{equation}

\noindent and for $Q(\lambda)\simeq\frac{\lambda}{k_{B,1}}\simeq\frac{\lambda}{\lambda_0}$. Therefore:

\begin{equation}
\begin{split}
    F(\lambda):=&\left(\frac{\lambda}{\lambda_0}\right)^3\left(1+\frac{\lambda_0}{k_{on}}\right)+\left(\frac{\lambda}{\lambda_0}\right)^2\left(\frac{P_{out}}{k_{on}}+K_D-1-\frac{\lambda_0}{k_{on}}\right)
    \\&+\left(\frac{\lambda}{\lambda_0}\right)\left(\frac{K_DP_{out}+P_{in}a_{ex}}{\lambda_0}-\frac{P_{out}}{k_{on}}-K_D\right)-K_D\frac{P_{out}}{\lambda_0}=0.
\end{split}
\end{equation}

\begin{center}
\begin{figure}[ht]
    \centering
    \begin{subfigure}{0.47\textwidth}
        \includegraphics[width=\textwidth]{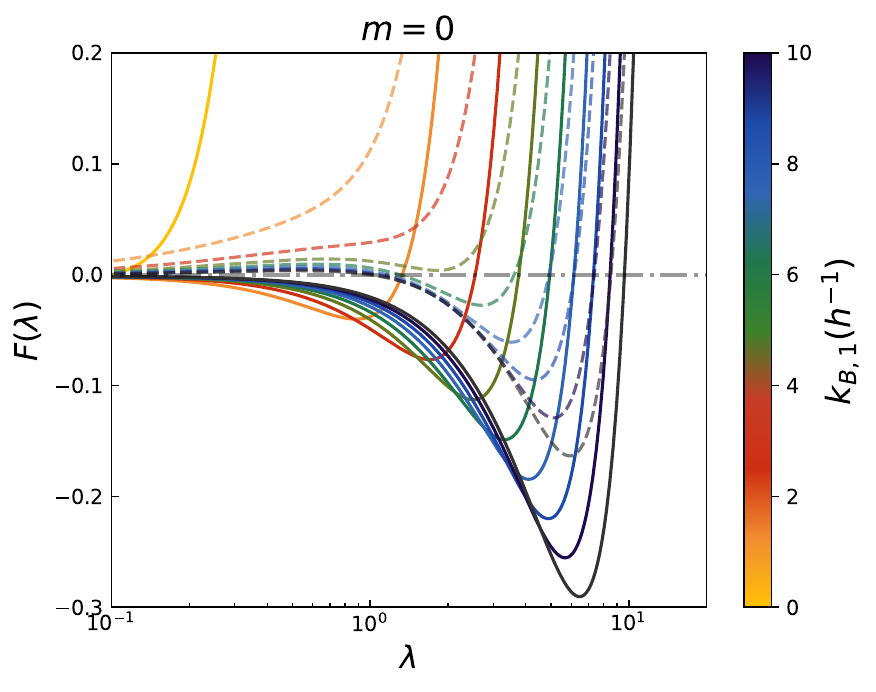}
        \caption{Exact self-consistent function defining the growth rate for $m=0$.}
        \label{fig:selfcons_m0_ex}
    \end{subfigure}
    \begin{subfigure}{0.47\textwidth}
        \includegraphics[width=\textwidth]{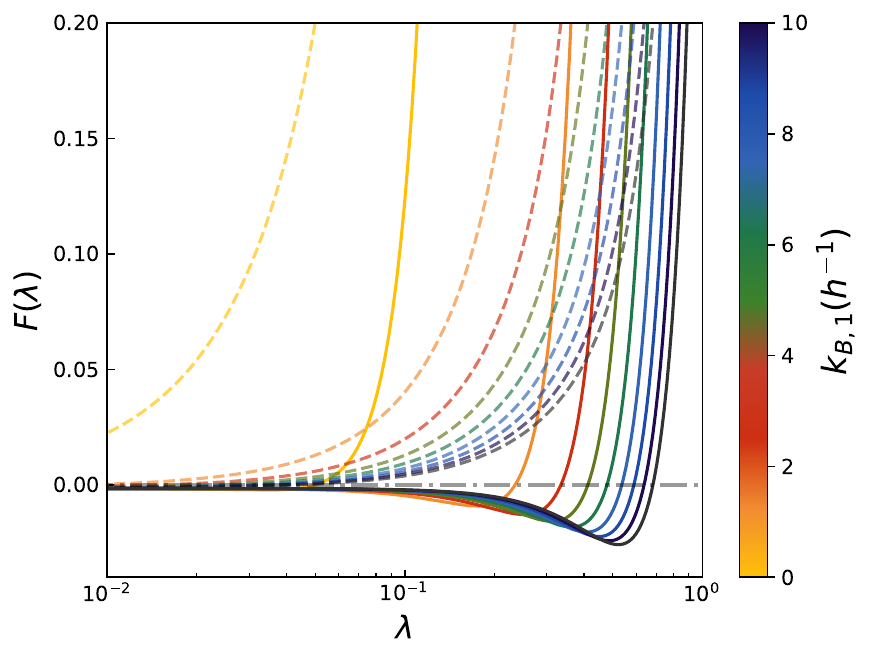}
        \caption{Exact self-consistent function defining the growth rate for $m=1$.}
        \label{fig:selfcons_m1_ex}
    \end{subfigure}
    \caption{Self-consistent function, the roots of which define the growth rate. The dotted lines represent the function with increasing values of $a_{ex}$.}
    \label{fig:selfcons}
\end{figure}
\end{center}

\noindent 
Let us define the parameter $m$ in such a way that the highest degree of the polynomial $Q(\lambda)$ is $m+1$, $m$ is also the number of limiting intermediate steps on Fig.1 of the main text. Thus in the present case since $Q(\lambda)$ is linear in $\lambda$, $m=0$.
In Fig.\ref{fig:selfcons}a, we plot the corresponding self-consistent function $F(\lambda)$, the roots of which correspond to the growth rates accessible to the system.

\noindent Increasing the abundance of external inhibitors modifies the curvature of the self-consistent function, in particular the concave part of the function vanishes above a given concentration of toxic agents. For small $m$, the minimum of the function can become positive and this will induce a discontinuity in the growth rate because of the concave part of the polynomial. For higher values of $m$, this effect is attenuated, which smooths the behaviour of the growth rate. We also recover different possible behaviours for the growth rate, in particular the reversible and irreversible limits. As discussed in the main text, Eq.\ref{eq:self_consistent_growthrate_general} has two solutions in the irreversible limit, leading to two separate branches of solutions for $\lambda$.


\subsubsection{Limiting intermediate steps}
\noindent Now, if we suppose that $m$ steps are considerably longer than the others, the growth rate of the system is $\lambda$ given by Eq.\ref{eq:self_consistent_growthrate_general} and $Q(\lambda)\simeq \left(\frac{\lambda}{\lambda_0}\right)^{m+1}$. Thus, the self-consistent equation can be written in terms of a function $F(\lambda)$:

\begin{equation}
\begin{split}
    F(\lambda) = &\left(\frac{\lambda}{\lambda_0}\right)^{2m+3}+\left(\frac{\lambda}{\lambda_0}\right)^{m+3}\frac{\lambda_0}{k_{on}}+\left(\frac{\lambda}{\lambda_0}\right)^{m+2}\left(\frac{P_{out}}{k_{on}}+K_D-1\right)
    \\&+\left(\frac{\lambda}{\lambda_0}\right)^{m+1}\frac{K_DP_{out}+P_{in}a_{ex}}{\lambda_0}-\left(\frac{\lambda}{\lambda_0}\right)^{2}\frac{\lambda_0}{k_{on}}-\left(\frac{\lambda}{\lambda_0}\right)\left(\frac{P_{out}}{k_{on}}+K_D\right)-K_D\frac{P_{out}}{\lambda_0}=0.
\end{split}
\label{poly_m_steps}
\end{equation}

\noindent This function is shown for the particular case of $m=1$ in Fig.\ref{fig:selfcons}b.

\medskip
\section{Experimental data and fitting procedure}

\subsection{List of compounds analyzed in this work}
\begin{figure*}
    \centering
    \begin{subfigure}{0.43\textwidth}
        \includegraphics[width=\textwidth]{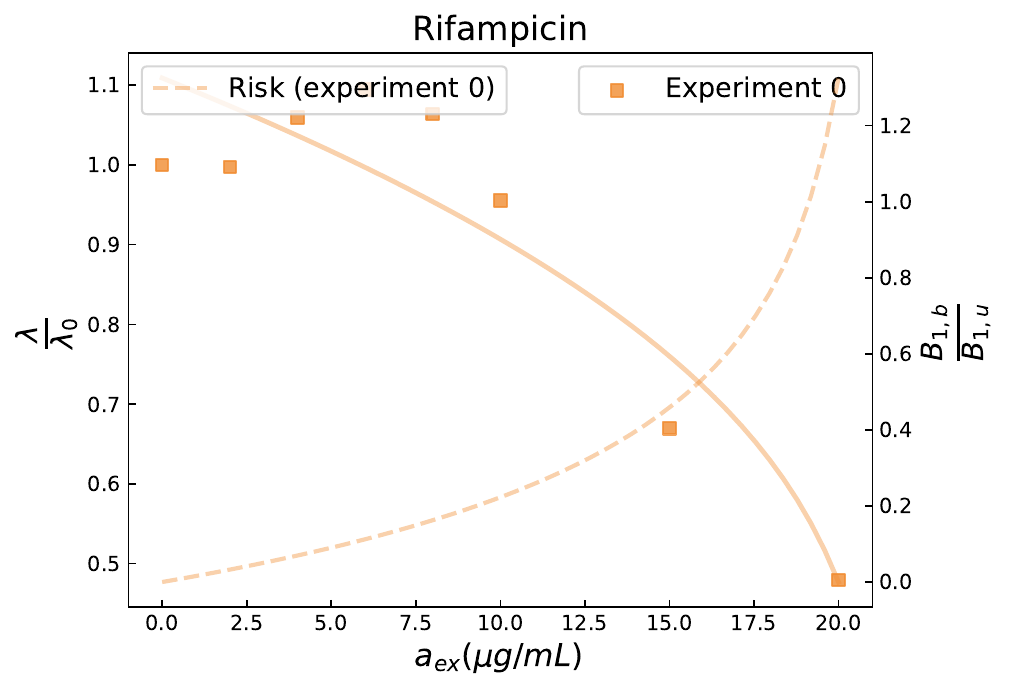}
        \caption{Rifampicin targets RNA-polymerase and inhibits RNA synthesis\cite{mcclure_mechanism_1978}.  Data from \cite{si_invariance_2017}.}
    \label{fig:rifampicin}
    \end{subfigure}
    \begin{subfigure}{0.39\textwidth}
        \includegraphics[width=\textwidth]{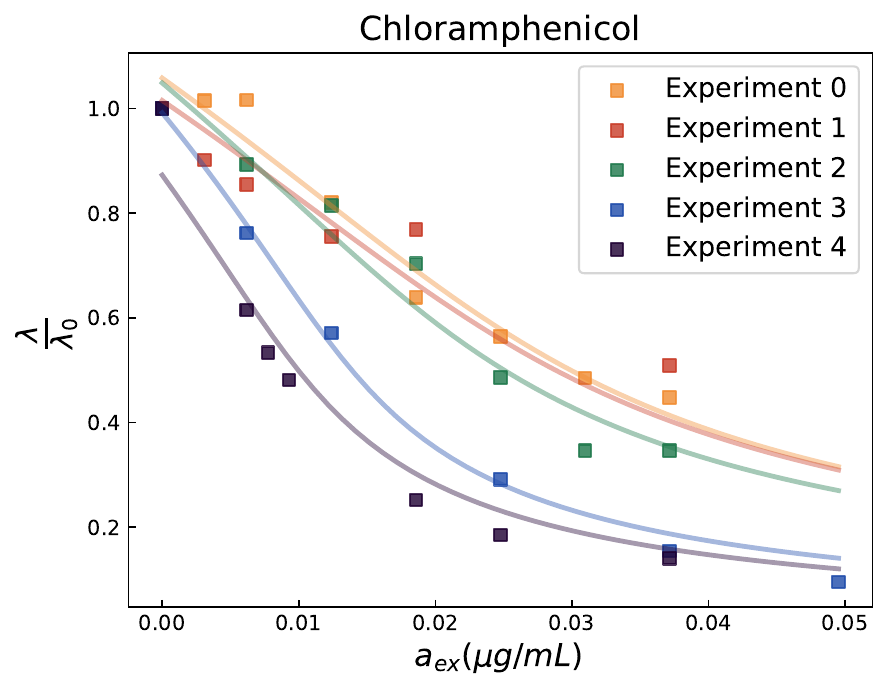}
        \caption{Chloramphenicol inhibits protein synthesis by binding to ribosomes \cite{contreras_binding_1974}. Data from \cite{si_invariance_2017}.}
    \label{fig:chloramphenicol}
    \end{subfigure}
    \begin{subfigure}{0.43\textwidth}
        \includegraphics[width=\textwidth]{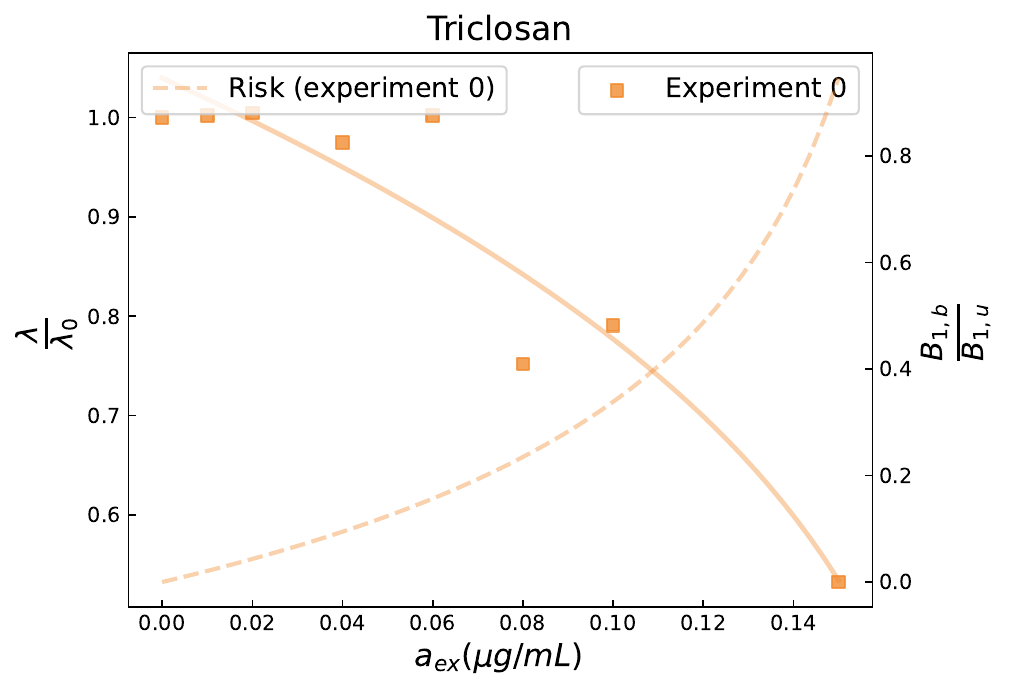}
        \caption{Triclosan targets the synthesis of fatty acids \cite{heath_mechanism_1999}. Data from \cite{si_invariance_2017}.}
    \label{fig:triclosan}
    \end{subfigure}
    \begin{subfigure}{0.43\textwidth}
        \includegraphics[width=\textwidth]{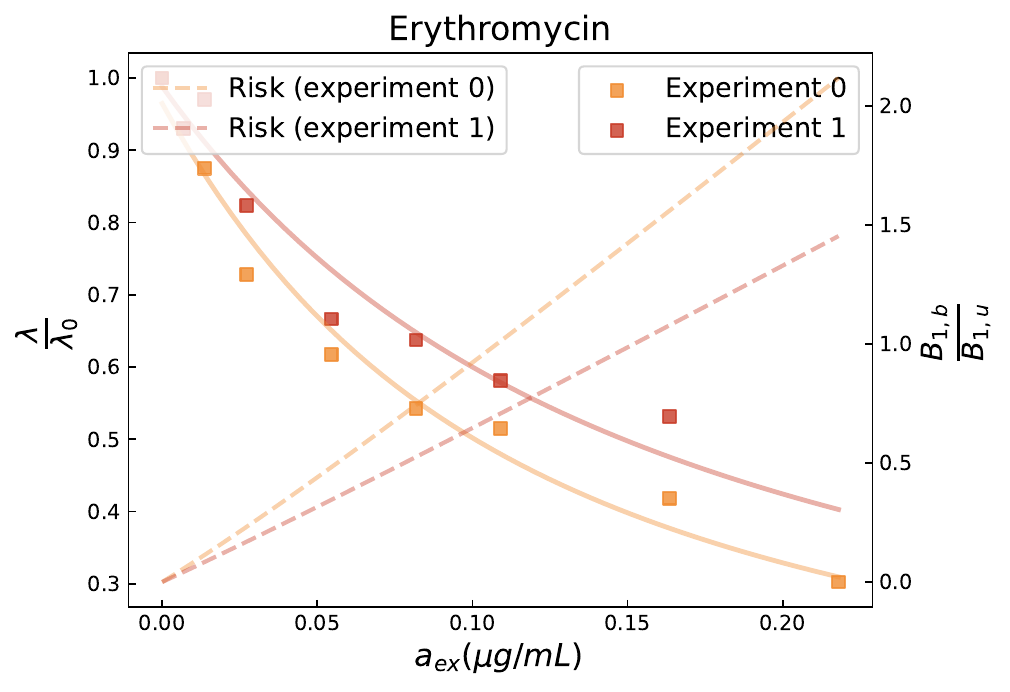}
        \caption{Erythromycin inhibits protein synthesis by binding to ribosomal proteins \cite{mazzei_chemistry_1993}. Data from \cite{si_invariance_2017}.}
    \label{fig:erythromycin}
    \end{subfigure}
    \begin{subfigure}{0.43\textwidth}
        \includegraphics[width=\textwidth]{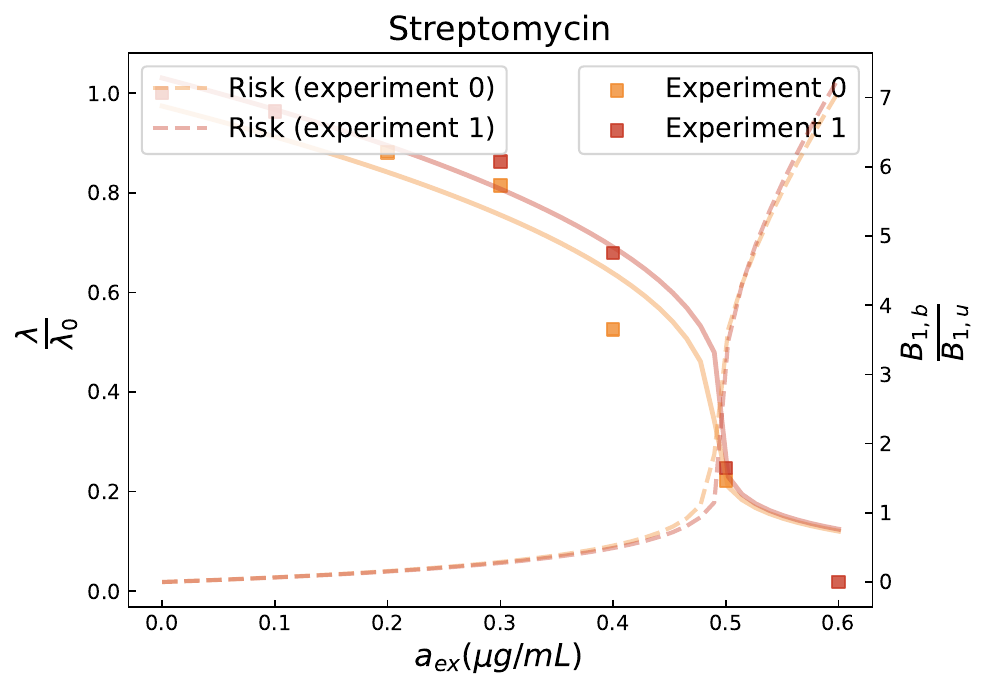}
        \caption{Streptomycin inhibits protein synthesis by binding to ribosomal proteins \cite{greulich_growth-dependent_2015}. Data from \cite{greulich_growth-dependent_2015}.}
    \label{fig:streptomycin}
    \end{subfigure}
    \begin{subfigure}{0.43\textwidth}
        \includegraphics[width=\textwidth]{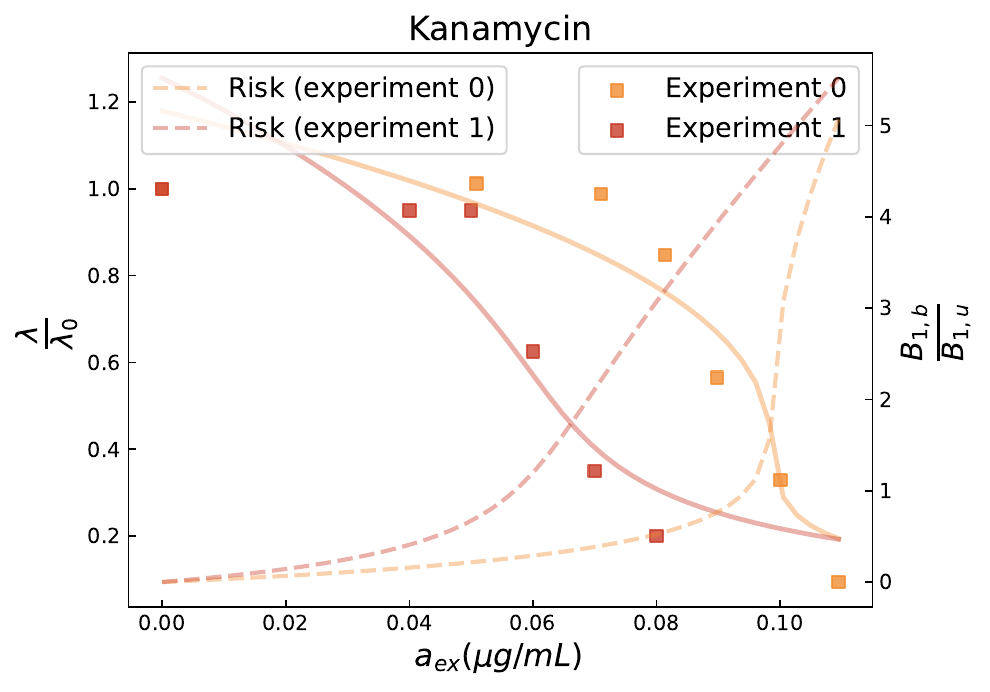}
        \caption{Kanamycin inhibits protein synthesis by binding to ribosomal proteins \cite{greulich_growth-dependent_2015}. Data from \cite{greulich_growth-dependent_2015}.}
    \label{fig:kanamycin}
    \end{subfigure}
    \caption{Comparison with experiments for various drugs. In solid lines, we show the growth rate as a function of the fraction of inhibitors. In dotted lines, we show a measure of the risk $\frac{B_{1,b} }{B_{1,u}}$. This measure compares the abundance of bound individuals $B_{1,b}$ to that of unbound operational individuals $B_{1,u}$ as in Eq \ref{eq:risk_simple}. For ribosome-targeting drugs, this corresponds to the fraction of bound ribosomes (inhibited) to unbound ribosomes (operating). Unbound ribosomes are indeed required for the vital functions of the cell whereas bound ribosomes are unable to synthesize proteins.}
    \label{fig:compare_drugs_app}
\end{figure*}

Chloramphenicol (Fig.\ref{fig:chloramphenicol}) inhibits ribosome production by binding to ribosomes (preventing them from transcribing new proteins). Its effect on growth laws has been studied \cite{greulich_growth-dependent_2015} as an example of bacteriostatic drug on E.Coli. Rifampicin (Fig.\ref{fig:rifampicin}) targets RNA-polymerase by binding to RNA-polymerase \cite{mcclure_mechanism_1978,campbell_structural_2001}(thus inhibiting the RNA-polymerase autocatalytic cycle discussed in \cite{roy_unifying_2021}). With our formalism, we also describe the effect of Triclosan (Fig.\ref{fig:triclosan}), Erythromycin (Fig.\ref{fig:erythromycin}), Streptomycin (Fig.\ref{fig:streptomycin}) and Kanamycin (Fig.\ref{fig:kanamycin}), which have different modes of action but are all bacteriostatic drugs against E.Coli. Kanamycin, Streptomycin, Chloramphenicol and Erythromycin target the ribosomal autocatalytic cycle at different stages and inhibit growth \cite{chaturvedi_protein_2016,lin_ribosome-targeting_2018,mondal_impact_2014,kohanski_mistranslation_2008}. Triclosan acts as a bacteriostatic by targeting the synthesis of fatty acids \cite{heath_mechanism_1999,mcmurry_triclosan_1998,escalada_triclosan_2005}, and thus affecting the building of bacterial membranes \cite{roy_unifying_2021}.

\subsection{Fitting procedure for the various antibiotics}

In order to recover the growth rate dependencies on drug concentration of Fig.\ref{fig:compare_drugs_app}, we fitted our expression Eq.\ref{eq:self_consistent_growthrate_general} with different sets of data, where $Q(\lambda)$ is given by Eq.\ref{eq:general_growthlaw}.

\begin{center}
\begin{table}[h]
    \centering\tiny{
    \begin{tabular}{|c|c|c|c|c|c|}
        \hline
         & $P_{in} (mL \cdot \mu g^{-1} \cdot h^{-1})$ & $P_{out} (h^{-1})$ & $k_{B,1} (h^{-1})$ & $k_{B,N+1} (h^{-1})$ & Experimental growth conditions
         \\
        \hline
         Triclosan & $2.85$ & $4.33$ & $1.28$ & $1.\times10^{-3}$ & MOPS glucose + 6 a. a.
         \\
        \hline
         Chloramphenicol (0) & $55.4$ & $44.4$ & $1.87$& $0.1$ & TSB (Tryptic soy broth)
         \\
        \hline
         Chloramphenicol (1) & $53.6$ & $44.4$ & $1.78$ & $0.1$ & TSB (Tryptic soy broth)
         \\
        \hline
         Chloramphenicol (2)  &$53.6$& $37.56$ & $1.52$ & $0.1$ & MOPS glucose synthetic rich
         \\
        \hline
         Chloramphenicol (3)  &$80.4$& $29.6$ & $1.66$ & $1.\times10^{-3}$ & Rich MOPS (0.2$\%$ glucose)
         \\
        \hline
         Chloramphenicol (4)  &$80.4$& $29.6$ & $1.18$ & $1.\times10^{-3}$ & Rich MOPS (0.2$\%$ glycerol)
         \\
        \hline
         Rifampicin &$2.20\times10^{-2}$& $3.46$ & $1.32$ & $1.0\times10^{-3}$ & MOPS glucose + 6 a. a.
         \\
        \hline
         Erythromycin (0) &$1.87\times10^2$& $8.9\times10^{2}$ & $1.16$ & $1.0\times10^{-3}$ & MOPS glucose + 6 a. a.
         \\
        \hline
         Erythromycin (1) &$1.35\times10^2$& $9.6\times10^2$ & $6.35\times10^{-1}$ & $1.\times10^{-1}$ & MOPS glycerol
         \\
        \hline
         Streptomycin (0) &$0.77$& $2.4$ & $1.04$ & $0.1$ & Rich MOPS (0.2$\%$ glycerol)
         \\
        \hline
         Streptomycin (1) &$0.69$& $1.85$ & $1.12$ & $9.9\times10^{-2}$ & MOPS (0.2$\%$ glucose, 0.2$\%$ casamino acids)
         \\
        \hline
         Kanamycin (0) &$3.60$& $1.97$ & $1.16$ & $8.0\times10^{-3}$ & MOPS (0.2$\%$ glycerol, 0.2$\%$ casamino acids)
         \\
        \hline
         Kanamycin (1) &$4.73$& $1.85$ & $1.09$ & $3.8\times10^{-3}$ & MOPS medium (0.2$\%$ glycerol)
         \\
        \hline
    \end{tabular}
    }
    \caption{Parameters estimated from the fitting procedure (using the package scipy.optimize)}
    \label{tab:results_fit}
\end{table}
\end{center}

We consider $N=6$ and separate the $N$ processes between fast and slow intermediary steps. For all antibiotics we assume that there is one no limiting step, to use the results of the main text. The steps are fast, and $(k_{B,n})_{2\leq n\leq6}$ are set to $10^5 h^{-1}$ (arbitrary high value compared to $\lambda$, in order to neglect those steps) so that $\lambda / k_{B,n} \ll 1$ for $n\geq 2$. For a given antibiotic, different experiments correspond to different growth conditions (\cite{greulich_growth-dependent_2015, si_invariance_2017}), that may affect the parameters of the model. As the number of free parameters is high, we constrained them in order to have biologically accurate values. From \cite{greulich_growth-dependent_2015, si_invariance_2017, koch_arthur_l_growth_1979}, we expect the basal growth rate $\lambda_0$ to be of order $1 h^{-1}$ (as measured in \cite{greulich_growth-dependent_2015}). The binding and unbinding rates, and the influx and outflux are expected to be faster, typically ranging between $1 h^{-1}$ and $1000 h^{-1}$ \cite{greulich_growth-dependent_2015, comby_mathematical_1989, svetlov_kinetics_2017}. From this considerations, we allow $k_{B,1}$ to vary between $0.4 h^{-1}$ and $4 h^{-1}$, $P_{out}$ to vary between $0 h^{-1}$ and $10^3 h^{-1}$ and $P_{in}$ to vary between $0 \mu g. mL^{-1}.h^{-1}$ and $10^3 \mu g. mL^{-1}.h^{-1}$ to capture the effects of reversibility. To reduce the number of free parameters, we set $K_D = k_{off}/k_{on}= 1/50 $ and $k_{off} = 5 h^{-1}$. And the deactivation rate $k_{B,N+1} \in [10^{-3} h^{-1}; 10^{-1} h^{-1}]$ is typically small compared to $\lambda$. From a biological point of view, as the different experiments used for one antibiotic correspond to various growth medium, we can consider that the reaction rates may vary from one experiment to the next, but we can assume that for a given antibiotic $P_{in}$ and $P_{out}$ weakly vary. By adding this constraint, there are $4$ parameters for each antibiotic but $P_{in}$ and $P_{out}$ cannot vary more than $20\%$ for different growth medium and a given antibiotic. In order to use concentrations in $\mu g / mL$ from the data in $\mu M$ for Chloramphenicol and Erythromycin we use molar masses ($323.132 g/mol$ for Chloramphenicol and $733.93 g/mol$ for Erythromycin).

\section{Complements}
\subsection{Combined effect of two antibiotics targeting different cycles}
\label{ap:cumulative_eff}

In order to produce Fig.6 of the main text, we solve the system for coupled autocatalytic cycles with two antibiotics obtained by keeping the $\min$ function in Eq.\ref{eq:starting_eqs} and using exponential solution (with the growth rate $\lambda$)

\begin{equation}
\begin{split}
    \lambda + \frac{1}{\tau_{life}}&=\frac{k_{B,2}}{\lambda + k_{B,2}+\frac{1}{\tau_{life(B)}}}\min (k_{B,1}\frac{B_{1,u}}{B_{tot}}, k_{C,1}\frac{C_{1,u}}{B_{tot}})
    \\ \lambda + \frac{1}{\tau_{life(C)}}&=\frac{k_{C,2}}{\lambda + k_{C,2}+\frac{1}{\tau_{life}}}\min (k_{B,1}\frac{B_{1,u}}{C_{tot}}, k_{C,1}\frac{C_{1,u}}{C_{tot}})
    \\ \frac{B_{1,b}}{B_{tot}} &= \frac{k_{on (B)}A_1}{\lambda + k_{off (B)} + \frac{1}{\tau_{life(B)}}}\frac{B_{1,u}}{B_{tot}}
    \\ \frac{C_{1,b}}{C_{tot}} &= \frac{k_{on (C)}A_2}{\lambda + k_{off (C)} + \frac{1}{\tau_{life(C)}}}\frac{C_{1,u}}{C_{tot}}
    \\ 1 - \frac{B_{1,u}}{B_{tot}} - \frac{B_{1,b}}{B_{tot}} & = \frac{k_{B,2}}{(\lambda + k_{B,2}+\frac{1}{\tau_{life(B)}})(\lambda + k_{B,3}+\frac{1}{\tau_{life(B)}})}\min(k_{B,1}\frac{B_{1,u}}{B_{tot}}, k_{C,1}\frac{C_{1,u}}{B_{tot}})
    \\ 1 - \frac{C_{1,u}}{C_{tot}} - \frac{C_{1,b}}{C_{tot}} & = \frac{k_{C,2}}{(\lambda + k_{C,2}+\frac{1}{\tau_{life(C)}})(\lambda + k_{C,3}+\frac{1}{\tau_{life(C)}})}\min(k_{B,1}\frac{B_{1,u}}{C_{tot}}, k_{C,1}\frac{C_{1,u}}{C_{tot}})
    \\ \frac{A_1}{B_{tot}} &= \frac{1}{\lambda + P_{out (B)} + k_{on}\frac{B_{1,u}}{B_{tot}}}\left(P_{in (B)} a_{ex,1} + k_{off} \frac{B_{1,b}}{B_{tot}}\right)
    \\ \frac{A_2}{C_{tot}} &= \frac{1}{\lambda + P_{out (C)} + k_{on}\frac{C_{1,u}}{C_{tot}}}\left(P_{in (C)} a_{ex,2} + k_{off} \frac{C_{1,b}}{C_{tot}}\right).
\end{split}
\end{equation}

The two first equations correspond to summing the equations for $B_{1,u}, B_{1,b}$ and $B_{3}$ (resp. for $C_{1,u}, C_{1,b}$ and $C_{3}$). The third and fourth equations are obtained from the equations on $B_{1,b}$ (resp. $C_{1,b}$). The fifth and sixth equations result from the equation on $B_2$ replaced in the equation for $B_3$. Finally, the two last equations follow from the equations $A_1$ and $A_2$.

\paragraph{Example with one building step in each cycle, one irreversible cycle and one reversible cycle} 

If the first cycle is reversible, we have on one hand
\begin{equation}
    \lambda = \frac{k_{B,1}}{1+\frac{K_{D(B)} P_{in(B)}}{P_{out(B)}}a_{ex,1}},
\end{equation}

and on the other hand, as the second cycle is irreversible
\begin{equation}
    \lambda = \frac{k_{C,1}}{2}\left(1+\sqrt{1-\frac{4P_{in(C)}a_{ex,2}}{k_{C,1}}}\right).
\end{equation}

For small concentrations of both antibiotics, we get a linear transition, which we observe on Fig.6 of the main text
\begin{equation}
    a_{ex,2} = \frac{k_{B,1} K_{D(B)}P_{in(B)}}{P_{in(C)}P_{out(B)}}a_{ex,1} + \frac{k_{C,1}-k_{B,1}}{P_{in(C)}},
\end{equation}

which simplifies into $a_{ex,2} = k_{B,1} a_{ex,1} K_{D(B)}/P_{out(B)} $ when the drugs have the same parameters for $P_{in}$ and $k_{B,1}$.

\paragraph{Example with one building step in each cycle, both reversible}

With two reversible cycles, for small concentrations of both antibiotics, in the particular case that both drugs have the same parameters, the transition occurs at $a_{ex,1}=a_{ex,2}$, which is observed on Fig.6 of the main text.

\subsection{Consequences of the inhibition of the first cycle on the second cycle}\label{ap:consequences}

\begin{figure}[!ht]
    \centering
    \includegraphics[width=0.55\textwidth]{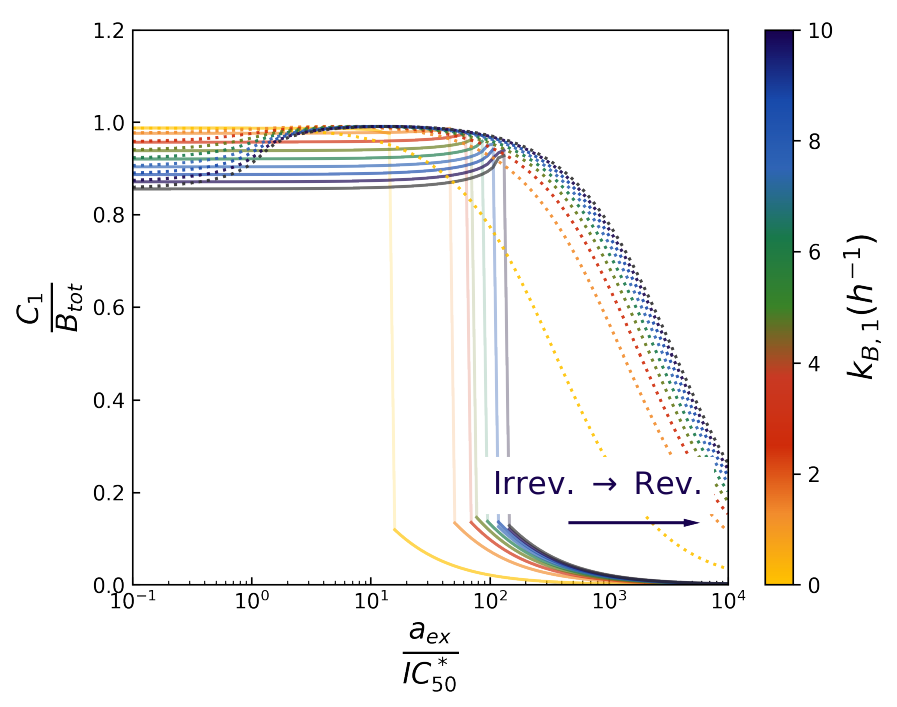}
    \caption{Fraction of autocatalysts in the second cycle when the first cycle is targeted by inhibitors.}
    \label{fig:second_cycle}
\end{figure}

To understand the effect of the $B$ cycle on the other one, the $C$ cycle in Fig.1 of the main text, we still assume $B$ species limiting, so the minimum function between $B_{1u}$ and $C_1$ in the equation for the production of $C_2$ gives $B_{1u}$. Now we focus on the equations for the $C$ species. Assuming again exponential growth with the same growth rate $\lambda$ in both cycles, we show the effect on $C_1$ on Fig.\ref{fig:second_cycle}. The increase in the relative amount $C_1/B_{tot}$ for intermediate values is only observed for $\tau_{life(C)}<\tau_{life(B)}$ and correspond to an accumulation of $C$ at long time. We recover the distinction between the reversible and irreversible cases. We also observe that there are regimes where $C_1$ increases with $a_{ex}$, which are obtained for $\tau_{life(C)}<\tau_{life(B)}$. 
Assuming again exponential growth with the same growth rate $\lambda$ in both cycles, we get:
\begin{equation}
\begin{split}
    \left(\lambda + \frac{1}{\tau_{life(C)}}\right)C_1 &= k_{C3}C_3-k_{C4}C_1
    \\ \left(\lambda + k_{C2}\right)C_2 &= k_{B1} B_{1,u}
    \\ \left(\lambda + \frac{1}{\tau_{life(C)}} + k_{C3}\right)C_3 &= k_{C4}C_1+k_{C2}C_2.
\end{split}
\end{equation}

\noindent Now if we introduce the total abundance of $C$, $C_{tot}=C_1 + C_3$:

\begin{equation}
\begin{split}
    \left(\lambda + \frac{1}{\tau_{life(C)}}\right)C_{tot} &= k_{C2}C_2
    \\ \left(\lambda + k_{C2}\right)C_2 &= k_{B1} Q(\lambda) B_{tot}
    \\ \left(\lambda + \frac{1}{\tau_{life(C)}} + k_{C3}\right)C_{tot} &= 
    \left(\lambda + \frac{1}{\tau_{life(C)}} + k_{C3}+k_{C4}\right)C_1+k_{C2}C_2.
\end{split}
\end{equation}

\noindent We can express everything in terms of $B_{tot}$:

\begin{equation}
\begin{split}
     C_{tot} &= \frac{k_{B,1}Q(\lambda)}{\left(\lambda+\frac{1}{\tau_{life(C)}}\right)\left(1+\frac{\lambda}{k_{C2}}\right)}B_{tot},
    \\ C_2 &= \frac{k_{B,1}Q(\lambda)}{\lambda + k_{C2}}B_{tot},
    \\ C_1 &= \frac{k_{C3}}{\lambda\hspace{-0.1cm}+\hspace{-0.1cm}\frac{1}{\tau_{life(C)}}+k_{C3}+k_{C4}}\frac{k_{B,1}Q(\lambda)}{\left(\lambda+\frac{1}{\tau_{life(C)}}\right)\hspace{-0.1cm}\left(1+\frac{\lambda}{k_{C2}}\right)}\hspace{-0.05cm}B_{tot}.
\end{split}
\end{equation}

\noindent From this we see that the second cycle is affected by the toxic agent via the growth rate. Note that here the difference in lifetimes matters, because as $\lambda \to 0$, we get (fast activation) 
\begin{equation}
\frac{C_1}{B_{tot}} \sim \frac{k_{C_3}}{\frac{1}{\tau_{life(C)}}+k_{C_3}+k_{C_4}}\frac{\tau_{life(C)}}{\tau_{life(B)}}\sim\frac{\tau_{life(C)}}{\tau_{life(B)}}.
\end{equation}


In addition, we observe that for small enough values of $a_{ex}$, the relative abundance of $C_1$ increases with $a_{ex}$. In this case, the slowing down of the first cycle does not affect strongly the second cycle. For large concentrations of antibiotics, the first cycle is frustrated and the second one becomes limited by the need for autocatalysts of type $B$, thus leading to lower relative abundances of $C_1$.

\subsection{Closed compartment and inhibiting waste}
\noindent For a closed compartment $P_{in}=P_{out}=0$ and waste $W$ produced at rate $k_w$, the equations are

\begin{equation}
\begin{split}
    \left(\lambda+k_{on}\frac{W}{B_{tot}}+k_{B4} + k_w\right) B_{1,u} &= k_{B3}B_3 + k_{off} B_{1,b}
    \\ \left(\lambda  + k_{off}\right)B_{1,b} &= k_{on} \frac{W}{B_{tot}} B_{1,u}
    \\ \left(\lambda + k_{B2}\right)B_2 &= k_{B1}B_{1,u}
    \\ \left(\lambda+k_{B3}\right)B_3 &= k_{B2}B_2 + k_{B4}B_{1,u}
    \\ \left(\lambda  + k_{on} \frac{B_{1,u}}{B_{tot}}\right)W &= k_{off} B_{1,b} + k_w B_{1,u}.
\end{split}
\end{equation}

Therefore, we can express $W/B_{tot}$ as a function of $Q(\lambda)$ and replace it in the second equation to find the risk
\begin{equation}
    \frac{B_{1,b}}{B_{1,u}}=\frac{k_{on}k_wQ(\lambda)}{\left(\lambda+k_{off}\right)\left(\lambda+k_{on}Q(\lambda)\frac{\lambda}{\lambda+k_{off}}\right)}.
\end{equation}

\end{document}